\documentclass[prd,preprint,superscriptaddress]{revtex4}
\usepackage{amsmath}
\usepackage{amssymb}
\usepackage{revsymb}
\usepackage{graphicx}

\newcommand{\be}{\begin{equation}}
\newcommand{\ee}{\end{equation}}
\newcommand{\ben}{\begin{eqnarray}}
\newcommand{\een}{\end{eqnarray}}

\newcommand{\lsim}{\lower.7ex\hbox{$\;\stackrel{\textstyle<}{\sim}\;$}}
\newcommand{\gsim}{\lower.7ex\hbox{$\;\stackrel{\textstyle>}{\sim}\;$}}

\begin{document}

\title{Late evolution of relic gravitational waves in coupled dark energy models}
\author{ Mar\'ia Luisa Sosa Almaz\'an and Germ\'an Izquierdo}
\address{  Facultad de Ciencias, Universidad Aut\'onoma del Estado de M\'exico, Toluca 5000, Instituto literario 100, Edo. Mex.,M\'exico.}
\email{gizquierdos@uaemex.mx}
\date{\today}
\begin{abstract}
 In this paper we study the late evolution of Relic Gravitational Waves in coupled dark energy models, where dark energy interacts with cold dark matter. Relic Gravitational Waves are second tensorial order perturbations of the Lemaitre-Friedman-Robertson-Walker metric and experiment an evolution ruled by the scale factor of the metric. We find that the amplitude of Relic Gravitational Waves is smaller in coupled dark energy models than in models with non interacting dark energy . We also find that the amplitude of the waves predicted by the models with coupling term proportional to the dark energy density is smaller than those of the models with coupling term proportional to dark matter density.
\end{abstract}

\maketitle

\section{Introduction}
Recent observational data indicates that the Universe is experimenting an accelerating expansion stage \cite{WMAP, PLANK}. The most favored model to explain the observations is the $\Lambda$CDM model, a Lemaitre-Friedman-Robertson-Walker (LFRW) universe with no curvature, and three dominant sources of energy density: barionic matter, cold dark matter and a cosmological constant \cite{rev}. On the other hand, the $\Lambda$CDM model presents the coincidence problem: why dark matter has an energy density comparable to that of the cosmological constant precisely today? (the former being a dynamical quantity that evolves through the history of the Universe and the latter being a constant). In an attempt to solve the coincidence problem, other models have been proposed \cite{rev}. In quintessence and phantom models, the role of the cosmological constant is played by the dark energy, a perfect fluid with equation of state $\rho_{\phi}= w p_{\phi}$,  where $-1 \gtrsim w$ and $w \lesssim -1$ respectively. For those models the dark energy density is also a dynamical function that evolves with time, and the coincidence problem is transformed in the fine tuning problem: the initial dark energy density and $w$ parameter must be tuned so that the dark energy density becomes of the same order than the dark matter density today.

In order to solve (or alleviate) the fine tuning problem, the coupled dark energy models (CDE) are introduced \cite{rapidc}. If the dark energy interacts with the dark matter, then their evolutions are not independent of each other. In fact they form a dynamical system and they have similar values today as the dynamical system evolve to a future attractor. Given that both sources are the "dark sector" of the Universe (in the sense that no direct observation of any of them have been recorded to date), a possible coupling between them must be not discarded beforehand. CDE models present a late evolution that differs from that of the $\Lambda$CDM model. Perturbations of the background metric evolve in a different way and should be addressed in order to test the validity of the CDE models  and/or to bound their free parameters. An example of this procedure can be found in \cite{ol}, where the density perturbations evolution is computed in different CDE models. The authors conclude that density perturbations present a damping in their amplitudes when they evolve in CDE scenarios instead of the non-coupled dark energy models. They also find a bound on the coupling parameter of the model. In \cite{He}, density perturbations evolution are also addressed in CDE models when the CDM perturbation experiments a collapse and clusters. The authors demonstrate that the energy density do not always fully cluster with the CDM, and that the cluster abundance count bounds the parameters of the CDE model considered.

Relic gravitational waves (RGWs) are primordial second tensorial order perturbations originated at the Big Bang. RGWs mechanism of amplification through the universe expansion is well-known \cite{Lifs, grish74}. RGWs evolution in non-coupled dark energy models has been addressed in the literature \cite{iz04}.  Obtaining data of RGWs spectrum would make possible to reconstruct the scale factor of the Universe. Even obtaining some observational bounds on the spectrum of RGWs would help to discard some dark energy models. Several efforts are currently being made in order to detect gravitational waves (e.g. ground based detectors LIGO, VIRGO, space based detector NGO, etc.). RGWs have a small energy density and its detection by the current experiments is doubtful, although depending on the inflationary models considered this view changes \cite{buo}. Also, 7-year WMAP data has been analyzed to bound the power spectrum of RGWs \cite{grish10}. Recently, the discovery of B mode polarization of the cosmic microwave background anisotropy has been announced \cite{BICEP}. The presence of RGWs at the last scattering surface would generate  the B mode polarization \cite{sel}. It is possible to reconstruct or at least to bind the RGWs spectrum through the observed data \cite{cheng}. In this work, we study the late evolution of RGWs in the CDE models. Given that the dark energy cannot be observed directly, theoretical bounds and/or indirect measurements (such as the bounds to the RGWs spectrum) are our only tools to understand the nature of it.

The plan of the article is the following. In section \ref{s2}, we briefly review the RGWs amplification mechanism. In section \ref{s3}, we address the dynamics of the CDE models. In section \ref{s4}, we discuss the free parameters of the model and we numerically evaluate the evolution of the RGWs amplitude for different choices of the parameters. In section \ref{s5}, we estimate the power spectrum of the RGWs of the previous section. Finally, in section \ref{s6}, we summarize the findings.

From now on, we assume units for which $c=\hbar=k_B=1$. As usual, a zero subindex refers to the current value of the corresponding quantity; likewise we normalize the scale factor of the metric by setting $a_{0} = 1$.

\section{RGW evolution from the Big Bang until the dust era}
\label{s2}

We consider a flat LFRW universe
\[
ds^{2}=-dt^{2}+a(t)^{2}\left[ dr^{2}+r^{2}d\Omega ^{2}\right] =a(\eta
)^{2}[-d\eta ^{2}+dr^{2}+r^{2}d\Omega ^{2}],
\]%
where $t$ and $\eta $ are, respectively, the cosmic and conformal time ($a(\eta )d\eta ={dt}$).

By introducing a small perturbation of the metric ($\overline{g}_{\alpha \beta}=g_{\alpha \beta}+h_{\alpha \beta}$,
$\left\vert h_{\alpha \beta}\right\vert \ll \left\vert g_{\alpha \beta}\right\vert $, $%
\alpha, \beta=0,1,2,3$) the perturbed Einstein equations follow. To linear order, the
transverse-traceless tensor solution which represents sourceless weak RGWs can be expressed as \cite{Lifs, grish74}%
\[
h_{i j}(\eta ,\overline{\mathbf{x}})=\int h_{i j}^{(\overline{k})}(\eta ,\overline{\mathbf{x}}%
)d^{3}\overline{k},
\]%
\begin{equation}
h_{i j}^{(\overline{k})}(\eta ,\overline{\mathbf{x}})=\frac{\mu (\eta )}{a(\eta )}%
G_{i j}(\overline{k},\overline{\mathbf{x}}),
\end{equation}%
where latin indices run from $1$ to $3$, and $\overline{k}$ is the commoving
wave vector. The functions $G_{i j}(\overline{k},\overline{\mathbf{x}})$ and $%
\mu (\eta )$ satisfy the equations
\begin{equation}
{G_{i }^{j}}_{;m }^{;m }=-k^{2}G_{i }^{j},\qquad {G_{i }^{j}}_{;j}=G_{i }^{i }=0,  \label{G}
\end{equation}%
\begin{equation}
\mu ^{\prime \prime }(\eta )+\left( k^{2}-\frac{a^{\prime \prime }(\eta )}{%
a(\eta )}\right) \mu (\eta )=0,  \label{eqmu}
\end{equation}%
where the prime indicates derivative with respect conformal time and $%
k=\left\vert \overline{k}\right\vert $ is the constant wave number related to the
physical wavelength and frequency by $k=2\pi a/\lambda =2\pi af=a\ \omega .$

The functions ${G_{i }^{j}}$ are combinations of $\exp (\pm i%
\mathbf{\overline{k}\overline{x}})$ which contain the two possible polarizations of the wave,
compatibles with the conditions (\ref{G}).

The equation (\ref{eqmu}) can be interpreted as an oscillator parametrically
excited by the term $a^{\prime \prime }/a$. When $k^{2}\gg \frac{a^{\prime
\prime }}{a}$, i.e., for high frequency waves, expression (\ref{eqmu})
becomes the equation of a harmonic oscillator whose solution is a free wave.
The amplitude of $h_{i j}^{(\overline{k})}(\eta ,\overline{\mathbf{x}})$ will decrease
adiabatically as $a^{-1}$ in an expanding universe. In the opposite regime,
when $k^{2}\ll \frac{a^{\prime \prime }}{a}$, the solution to (\ref{eqmu})
is a lineal combination of $\mu _{1}\propto a(\eta )$ {and }$\mu _{2}\propto
a(\eta )\int d\eta \ a^{-2}$. In an expanding universe $\mu _{1}$ grows
faster than $\mu _{2}$ and will soon dominate. Accordingly, the amplitude
of $h_{i j}^{(\overline{k})}(\eta ,\overline{\mathbf{x}})$ will remain constant so long
as the condition $k^{2}\ll \frac{a^{\prime \prime }}{a}$ is satisfied. When
it is no longer satisfied, the wave will have an amplitude greater than it
would in the adiabatic behavior. This phenomenon is known as \textquotedblleft parametric
amplification\textquotedblright\ of relic gravitational waves \cite{grish93,
grish74}.

For power law expansion $a\propto \eta ^{l}$ ($l=-1,1,2$ for inflationary,
radiation dominated and dust dominated universes, respectively) the solution
to equation (\ref{eqmu}) is
\[
\mu (\eta )=(k\eta )^{\frac{1}{2}}\left( K_{1}J_{l-\frac{1}{2}}(k\eta
)+K_{2}J_{-\left( l-\frac{1}{2}\right) }(k\eta )\right) ,
\]%
where $J_{l-\frac{1}{2}}(k\eta )$ , $J_{-\left( l-\frac{1}{2}\right) }(k\eta
)$ are Bessel functions of the first kind and $K_{1,2}$ are integration
constants.

We assume now that the universe experiments and inflationary stage of evolution followed by a radiation dominated stage and a dust stage \cite{grish93}. Transitions
between successive eras are assumed instantaneous. This approach is known as sudden transition approximation, and it is a reasonable approximation when the transition time span between the different stages is much lower than the period of the RGWs considered. The scale factor, then, is%
\begin{equation}
a(\eta )=\left\{
\begin{array}{c}
-\frac{1}{H_{1}\eta }\qquad (-\infty <\eta <\eta _{1}<0), \\
\frac{1}{H_{1}\eta _{1}^{2}}(\eta -2\eta _{1})\qquad (\eta _{1}<\eta <\eta
_{2}), \\
\frac{1}{4H_{1}\eta _{1}^{2}}\frac{(\eta +\eta _{2}-4\eta _{1})^{2}}{\eta
_{2}-2\eta _{1}}{\qquad }(\eta _{2}<\eta )%
\end{array}%
\right.  \label{sclfac1}
\end{equation}%
where the subindexes $1,2$ correspond to the sudden transitions from
inflation to radiation era and  from radiation to dust era, respectively, and $H_{1}$
represents the Hubble factor at the end of the inflationary era.

The solution to the equation (\ref{eqmu}) for each era is
\begin{eqnarray}
\mu _{I}(\eta ) &=&C_{I}\left[ \cos (k\eta +\phi _{I})-\frac{1}{k\eta }\sin
(k\eta +\phi _{I})\right] \qquad \text{(inflationary era)}  \label{muinfl} \\
\mu _{R}(\eta ) &=&C_{R}\sin (k\eta _{R}+\phi _{r})\qquad \qquad \qquad
\qquad \qquad \qquad \text{(radiation era)} \\
\mu _{D}(\eta ) &=&C_{D}\left[ \cos (k\eta _{D}+\phi _{D})-\frac{1}{k\eta
_{D}}\sin (k\eta _{D}+\phi _{D})\right] \qquad \text{(dust era)},
\label{mudust}
\end{eqnarray}
where $C_{I,R,D}$, $\phi _{I,R,D}$ are constants of integration, $\eta
_{R}=\eta -2\eta _{1}$ and $\eta _{D}=\eta +\eta _{2}-4\eta _{1}$ .

It is possible to express $C_{R}$, $\phi _{R}$ \ and $C_{D}$, $\phi _{D}$ in
terms of $C_{I}$, $\phi _{I}$ and $C_{R},$ $\phi _{R}$ respectively as $\mu
(\eta )$ must be continuous at the transition times $\eta =\eta _{1}$ and $%
\eta =\eta _{2}$. Averaging the solution over the initial phase $\phi _{I}$
the amplification factor is found to be

\[
R(k)=\frac{C_{D}}{C_{I}}\sim \left\{
\begin{array}{c}
1\qquad (k\gg -1/\eta _{1}), \\
k^{-2}\qquad (-1/\eta _{1}\gg k\gg 1/(\eta _{D2})), \\
k^{-3}\qquad (1/(\eta _{D2})\gg k).%
\end{array}%
\right.
\]

The evolution of RGW from the instant $\eta_2$ up to the present day in the standard cosmological model is addressed in \cite{iz04}. As the universe experiments a late accelerated stage the potential term $a''/a$ becomes an increasing function of $\eta$. Consequently,  some waves that already where in the  adiabatic regime reenter the $k^2<<a''/a$ regime, and cease to contribute to the physical power spectrum of RGWs.

In the next section, we consider a different scenario where a coupled dark energy (CDE) stage follows the radiation stage.

\section{RGW evolution from the dust era until the present day}
\label{s3}
\subsection{CDE models and universe expansion}
CDE energy models are an alternative to the $\Lambda$CDM model in an attempt to solve the coincidence problem \cite{rapidc}. Those models assume that after the radiation stage the universe expansion is dominated by a mixture of three dominant energy density sources: Barionic matter $\rho_{b}$, Cold Dark Matter (CDM) $\rho_{c}$ and Coupled Dark Energy $\rho_{\phi}$. CDE interacts with CDM through a positive interaction term $Q$ \cite{db} and, consequently, the energy densities evolve as

\begin{eqnarray}
\dot{\rho}_{b}\, &+&\, 3H \rho_{b} = 0 \, , \nonumber \\
\dot{\rho}_{c}\, &+& \, 3 H \rho_{c} = Q \, , \nonumber \\
\dot{\rho}_{\phi}\, &+& \, 3 H (1+w) \rho_{\phi} = - Q \, ,
\label{continuity}
\end{eqnarray}
where $w$ is the CDE adiabatic coefficient $p_{\phi}=w\rho_{\phi}$ and $H$ is the Hubble factor

\be
H^2=\left(\frac{\dot{a}}{a}\right)^2=\frac{8 \pi G}{3} (\rho_{b}+\rho_{c}+\rho_{\phi}).
\label{H}
\ee

In view of the above equations the coupling must be a function of
$H \rho_{c}$ and/or $H \rho_{\phi}$ , with a small proportionality constant \cite{iz10}. We will consider for simplicity two different coupling terms
\ben
Q_1 = \alpha H \rho_{c}\, , \\ Q_2 = \alpha H \rho_{\phi} \, ,
\label{int}
\een
where $\alpha$ is an adimensional positive coupling constant. At this point we can solve the evolution of the universe in terms of the free parameters of the model $w$ and $\alpha$ and the present day values of the densities $\rho_{b0}$, $\rho_{c0}$ and $\rho_{\phi 0}$.
\begin{itemize}

\item{Interaction proportional to $\rho_{c}$}

By plugging $Q_1$ into (\ref{continuity}),
assuming $w$ is constant, and integrating we get the energy densities dependence on the scale factor
\begin{eqnarray}
\rho_{b}&=& \rho_{b0} \, a^{-3} \, , \nonumber \\
\rho_{c} &=& \rho_{c0}\, a^{-3+ \alpha} \, , \nonumber \\
\rho_{\phi} &=& \rho_{\phi0}\, a^{-3(1+w)}\, + \, \rho_{c0} \,
\frac{\alpha}{3 w+\alpha} \, \left[ a^{-3(1+w)}-
a^{-3+\alpha} \right].
\label{rho(a)2}
\end{eqnarray}

\item{Interaction proportional to $\rho_{\phi}$}

Using $Q_2$ into (\ref{continuity}), we obtain
\ben
\rho_{b}&=& \rho_{b0} \, a^{-3} \, , \nonumber \\
\rho_{c}&=& \rho_{c0}\, a^{-3}+\frac{\alpha}{3 w+\alpha}\,
\rho_{\phi0} \,
a^{-3}\left[1-a^{-3w-\alpha}\right] \, , \nonumber \\
\rho_{\phi}&=&\rho_{\phi0} \, a^{-3(1+w)-\alpha} \, .
\label{rho(a)1} \een

\end{itemize}

In each case the Hubble factor is a function of the scale factor $a$ which can be obtained from the above solutions and (\ref{H}).

\subsection{RGW late evolution}

The amplitude of RGW, $\mu(\eta)$, evolves with conformal time as (\ref{eqmu}). The dynamics of the  LFRW universe affects the amplitude evolution through the potential term $a''/a$.

At this time it is convenient to rewrite equation (\ref{eqmu}) in terms of the scale factor $a$. From the definition of the conformal time and the Hubble factor $a d\eta= dt=da/(H a) $. Thus, $'=d/d\eta=a^2 H d/da$. The potential $a''/a$ can be expressed as
\be
\frac{a''}{a}(a)=2a^2 H^2(a)+a^3 H(a) \frac{dH}{da}(a).
\label{numpot}
\ee

The term $\mu''(\eta)$ of equation (\ref{eqmu}) is transformed to
\be
\mu''(a)= a^4 H^2(a) \frac{d^2 \mu}{da^2}(a)+\left( 2a^3 H^2(a)+a^4 H(a) \frac{dH}{da}(a) \right)\frac{d \mu}{da}(a).
\label{nummu''}
\ee

Note that while integrating equation (\ref{eqmu}) in terms of the conformal time $\eta$ the only term to consider is $a''/a$. But when integrating in terms of the scale factor $a$, an additional term proportional to $d \mu/da$ appears. We can now make use of the energy densities obtained for each interaction to evaluate the Hubble factor $H(a)$  and its derivative respect to $a$ ($dH/da$) and, eventually, to solve equation (\ref{eqmu}) by numerical methods.  We discuss the results obtained in the following section.

\section{Numerical results and discussion}
\label{s4}

First we have a look to the constants appearing in equations (\ref{rho(a)1}) and (\ref{rho(a)2}), the free parameters of the model ($\alpha$ and $w$), the initial conditions and the wave number $k$.

\begin{itemize}

\item{Constants $\rho_{b0}$, $\rho_{c0}$ and $\rho_{\phi 0}$}

To fix the present day energy densities, it is convenient to set a scale over the Hubble factor. We will asume from now on that $H_0=1$ (i.e. that
$(8 \pi G/3)(\rho_{b0}+\rho_{c0}+\rho_{\phi 0})=1$).

Then, it is straightforward that
\be
H^2(a)=\Omega_{b}(a)+ \Omega_{c}(a)+\Omega_{\phi }(a),
\label{hubfac}
\ee
where $\Omega_i(a)=\rho_i(a)/(\rho_{b0}+\rho_{c0}+\rho_{\phi 0})$.
Observational data define the present day values of $\Omega$ functions at $a=a_0=1$. WMAP data set those values to $\Omega_{b0}=0.04$, $\Omega_{c0}=0.24 $ and $\Omega_{\phi 0}=0.72$ \cite{WMAP}. The more recent PLANK observations set them to $\Omega_{b0}=0.05$, $\Omega_{c0}=0.26 $ and $\Omega_{\phi 0}=0.69$ \cite{PLANK}.

The CDE models given by equations (\ref{int}) have been tested to check whether the density fluctuations, BAO and supernovas data are in good agreement with those observed or not, by assuming the WMAP data (\cite{ol}). Those models allow the observed density fluctuations when $0 \leq \alpha \leq 0.1$. In \cite{March}, authors analyze  interaction $Q_1$ with negative values of the coupling parameter and the PLANCK data concluding that the observed density fluctuations allows an interaction parameter of the order of $-0.49$ to a $68\%$ confidence level. In this work we assume the WMAP values in our computations, as our coupling parameter is positive defined.

\item{Free parameters $\alpha$ and $w$}

The free parameters of the model are $\alpha$ and $w$. The first is related to the coupling term $Q$ and the second is related to the nature of dark energy.

When the parameter $\alpha=0$, dark energy is not coupled to CDM. On the other hand, for $\alpha>0.1$, the structure formation in the CDE model will differ from the predicted by the CMB data. The difference is due to the presence of too much dark energy at the decoupling instant \cite{ol}.

For our computation, we will assume that $0<\alpha\leq 0.1$. We choose three values in order to plot the $\mu$ function: $\alpha= 0.001, 0.01, 0.1$. The fist value represents an scenario where the term $a''/a$ do not differ considerably from the non-coupled dark energy model. The late value represents an scenario with a highly coupled dark energy and a $a''/a$ term different from the non-coupled dark energy scenario.

For the adiabatic parameter $w$ of dark energy, observational data from Barionic Acoustic Oscillations, Supernova Survey and WMAP data of CMB conclude that $w\sim -1$ ($w=-1$ being the cosmological constant model) \cite{rev}, but both quintessence ($w>-1$) and phantom $ w<-1$ models of dark energy are allowed. To numerically obtain and plot $\mu$, we will assume three possible values: $w=-0.9,-1,-1.1$.

\item{Initial conditions and RGW wave number $k$}

As it is mentioned early, the RGW amplitude $\mu(a)$ must be a continuous function. This fact allows to obtain a relation between the different solutions at different stages of expansion in the sudden transition approach.

In our scenario, the CDE stage of expansion represents a smooth evolution from a universe dominated by CDM  and barionic matter when $a \rightarrow 10^{-4}$ until the present day. In fact, depending on the value of the parameters $\alpha$ and $w$, the CDE universe will evolve as a dynamical system to an attractor of the system in which the ratio $\rho_{\phi}/\rho_c$ is constant \cite{rapidc}.

Equation (\ref{eqmu}) is a second order differential equation, and we need initial conditions for $\mu$ and $d\mu/da$ in order to solve it. When $a \rightarrow 10^{-4}$, the Hubble factor is approximately $H\simeq\sqrt{\Omega_{b0}+\Omega_{c0}}a^{-3/2}$. Thus, the conformal time can be integrated
\be
\int^\eta_{\eta_2} d\eta=\int^a_{10^{-4}} \frac{da}{a^2H} \Rightarrow a=(10^{-2}+\frac{1}{2}\sqrt{\Omega_{b0}+\Omega_{c0}}(\eta-\eta_2))^2.
\ee

From the scale factor of above and (\ref{eqmu}) in the limit $a \rightarrow 10^{-4}$, it is straightforward that the RGW amplitude will tend to the dust solution $\mu_D$ from (\ref{mudust}) with
\be
\eta _{D}=\frac{2}{\sqrt{\Omega_{b0}+\Omega_{c0}}}\left(10^{-2}+
\frac{1}{2}\sqrt{\Omega_{b0}+\Omega_{c0}}(\eta-\eta_2)\right)= \frac{2}{\sqrt{\Omega_{b0}+\Omega_{c0}}} a^{1/2},
\label{incond}
\ee
 Thus, it is very reasonable to use as initial conditions for the RGW amplitude
\ben
\mu(a=10^{-4})&=& \mu _{D}(\eta_D(a=10^{-4})),\nonumber \\
\frac{d \mu}{da}(a=10^{-4})&=& \frac{d \mu_D}{da}(\eta_D(a=10^{-4})).
\label{incond}
\een
The constants $C_D$ and $\phi_D$ connect the $\mu$ solution in the CDE era to the solutions in previous eras. For the sake of simplicity, we will assume that $\phi_D=0$. The constant $\phi _{D}$ is an initial phase and it is not important to the general evolution of the amplitude $\mu$. We will also set $C_D=1$ as a scale over the amplitude of $\mu$.

Concerning the wave number of the RGW, $k$, appearing in (\ref{eqmu}) and  (\ref{incond}), the reader should note that $\eta_D(a=10^{-4})=2\cdot10^{-2}/(\sqrt{\Omega_{b0}+\Omega_{c0}})$. The argument in the initial condition (\ref{mudust}) reads $k\eta_D(a=10^{-4})$. For the numerical work, it is convenient to define an adimensional wave number $\textrm{k}=k/(\sqrt{\Omega_{b0}+\Omega_{c0}}H_0)$, where we have momentarily recovered $H_0$ to stress the adimensionality of $\textrm{k}$, and the argument in the initial conditions is just $2\cdot10^{-2}\textrm{k}$. We will also choose three values of the wave number $\textrm{k}$ to represent the three regimes of the wave equation: $\textrm{k}_1=10^{3/2}$, for the free wave regime of equation (\ref{eqmu}) ($\textrm{k}^2 >> a''/a$); $\textrm{k}_2=10^{1/2}$ for the $\textrm{k}^2$ of the same order os magnitude than $a''/a$ regime, and $\textrm{k}_3=10^{-1/2}$ for the parametric regime ($\textrm{k}^2 << a''/a$).
\end{itemize}

At this point, we are in position to numerically solve equation (\ref{eqmu}), with the potential (\ref{numpot}) and $\mu''$ defined by (\ref{nummu''}), from an initial  $a=10^{-4}$ up to the present day value $a_0=1$ and considering the initial conditions (\ref{incond}).

\begin{figure}[tbp]
\includegraphics*[scale=0.4]{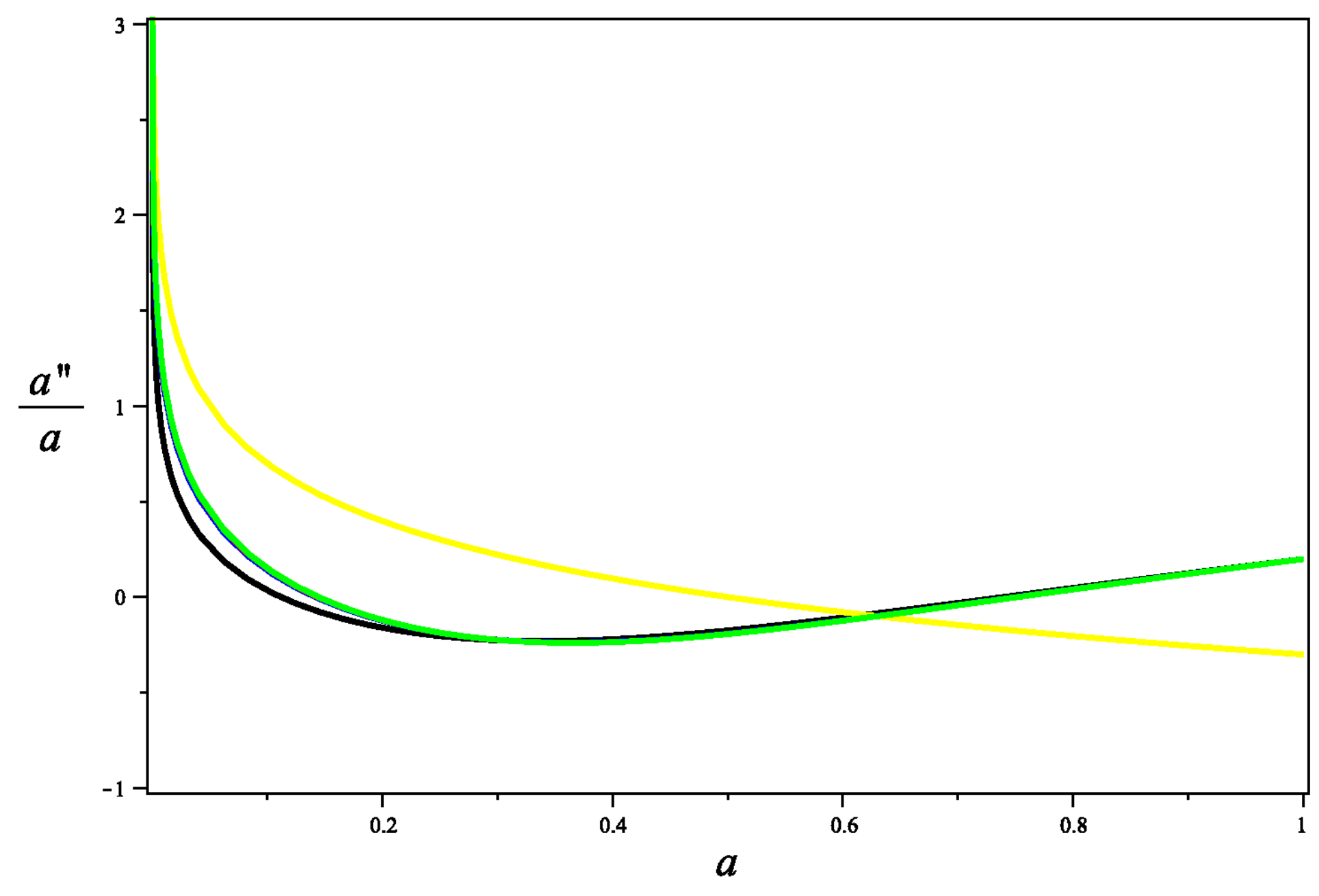}
\includegraphics*[scale=0.4]{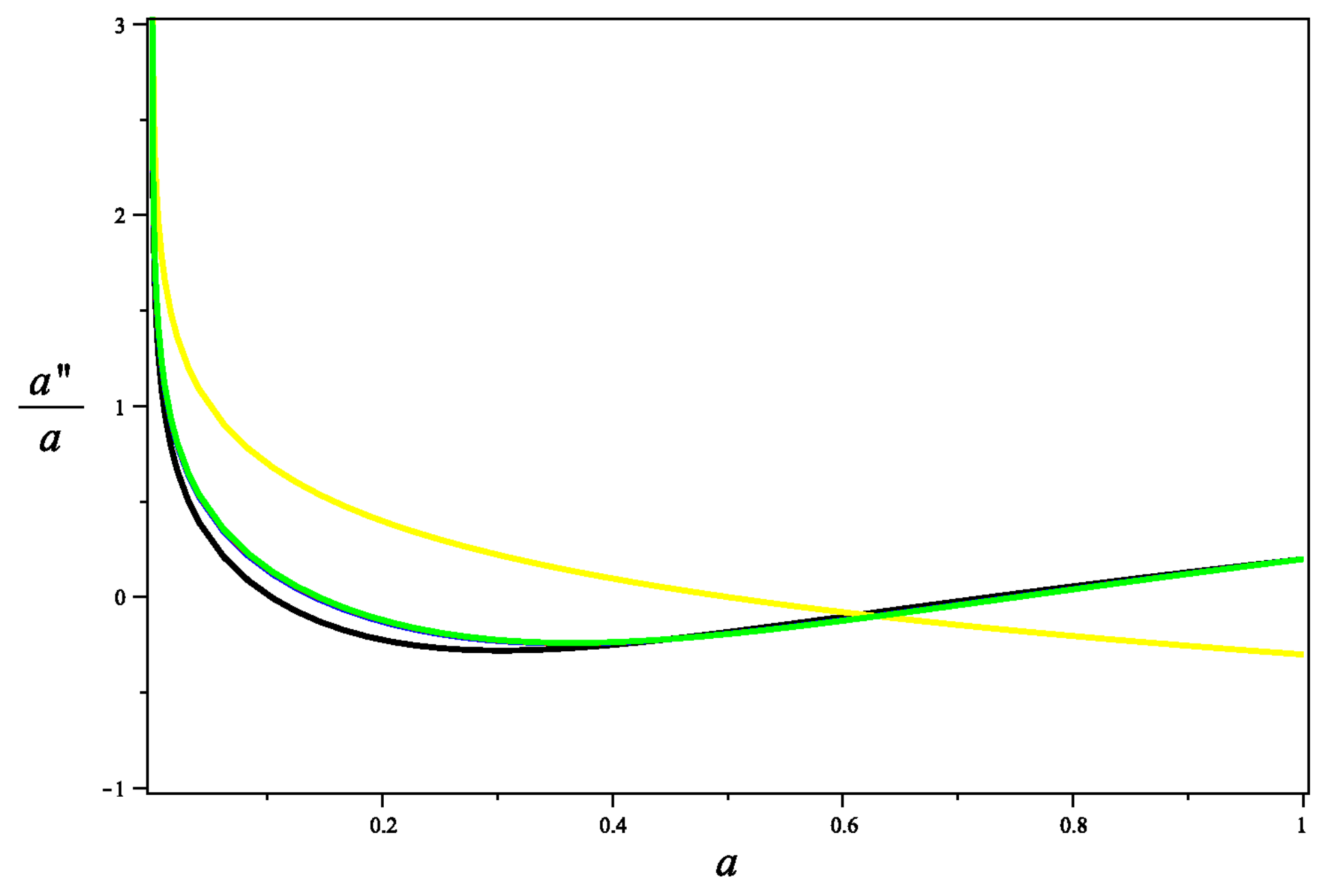}

\caption{Potential $a''/a$ vs. $a$ for different choices of the parameters $\alpha$, $w$ and. Left panel represents the interaction $Q_1$ and right panel the interaction $Q_2$.} \label{fig1}
\end{figure}

Once we fix the value of the model parameters $\alpha$ and $w$, we plot the potential term $a''/a$ for every interaction in a semi-logarithmic plot in terms of the scale factor $a$. Left panel of Figure \ref{fig1} represents the potential term for the interaction $Q_1$ with $w=-1$ and different values of $\alpha$, while right panel represents the potential for the interaction $Q_2$ and the same choice of parameters. In both panels, the black line represents the $\alpha=0.1$ scenario, grey line is the superposition of $\alpha=0.01$, $\alpha=0.001$ and $\alpha=0.0$ lines. Finally the  lighter line represents the potential term of a dust dominated universe (with no dark energy).

From the Figure \ref{fig1}, the reader may observe that there is a slight difference between the potential of the $\alpha=0.1$ and the one of the non coupled model $\alpha=0.0$, while the potential of the $\alpha=0.01$ and $\alpha=0.001$ models are undistinguishable to bare eye from the latter. There is also a difference between the potential of both $\alpha=0.1$ models for the different interactions. Comparing the scenarios with CDE with the scenario with no dark energy (dust scenario), we can state that the potential in the latter decreases at a slower rate. The former scenarios also present a region where the potential term grows with the scale factor (related to the acceleration of the expansion of the universe and the evolution of the event horizon \cite{iz04}) while in the latter scenario the potential is an ever decreasing function of $a$. While the shape of the potential term $a''/a$ is determinant to the evolution of the function $\mu$, we cannot conclude that $\alpha=0.0$, $\alpha=0.01$ and $\alpha=0.01$ models will have the same solution, as other terms have been introduced to equation (\ref{eqmu}) when we changed the variable of integration from $\eta$ to $a$.

We next plot the numerical solutions for different choices of $\alpha$, $w$. In the following figures the potential term $a''/a$ vs. $a$ is represented by the thick black line and its semi-logarithmic scale is shown in the left vertical axis. In the same scale as $a''/a$, we have represented the three values of $\textrm{k}^2$ considered.

We have plotted the numerical solutions $\mu$ {\it vs.} $a$ as the grey lines that oscillate around its respective $\textrm{k}^2$ lines, the scale for $\mu$ can be found in the right vertical axis (in this scale the initial amplitude $C_D$ is defined as 1). We have also plotted the initial condition $\mu_D$ from equation (\ref{mudust}) for each $\textrm{k}$ as the lighter line.

\begin{figure}[tbp]
\includegraphics*[scale=0.4]{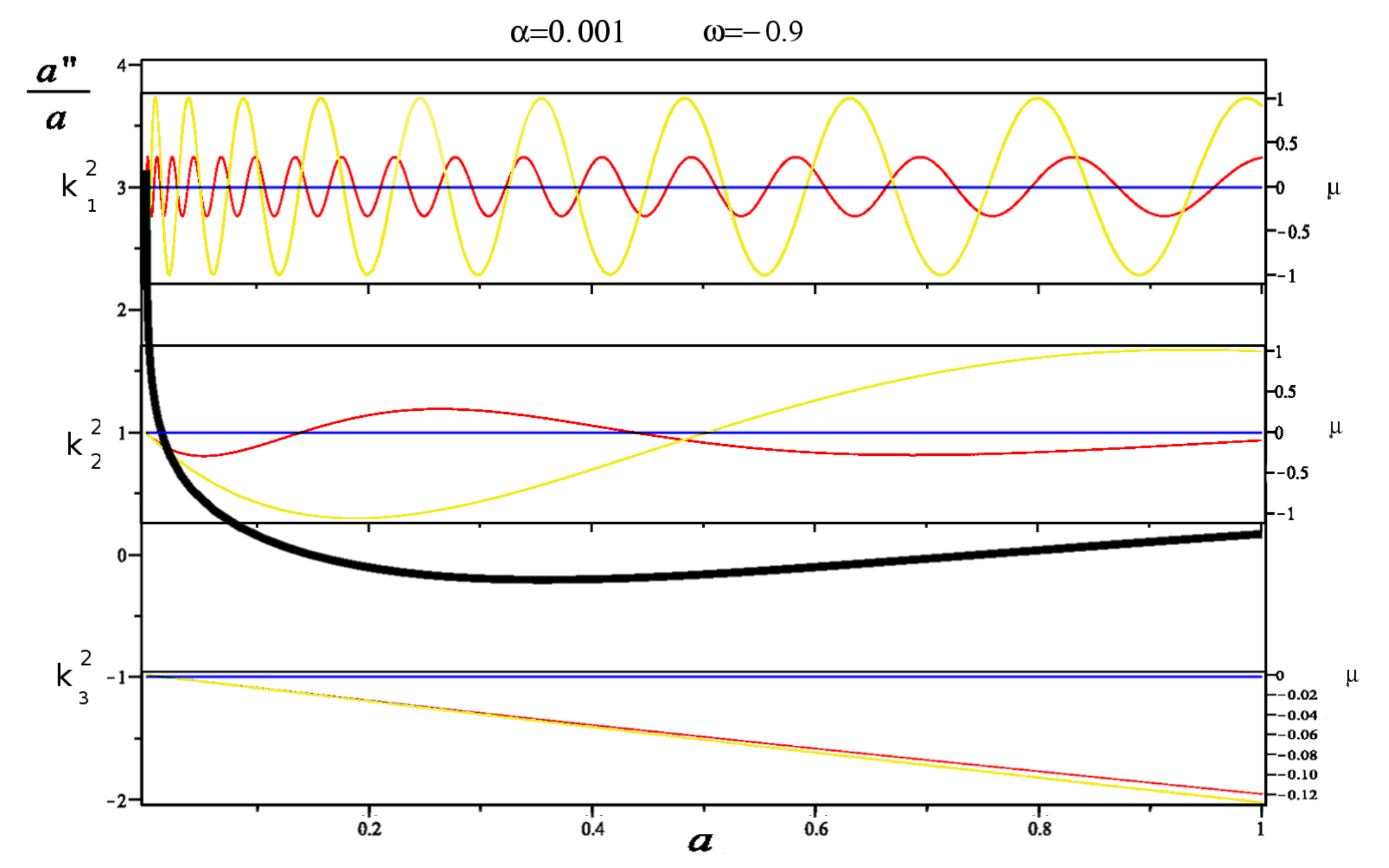}
\includegraphics*[scale=0.4]{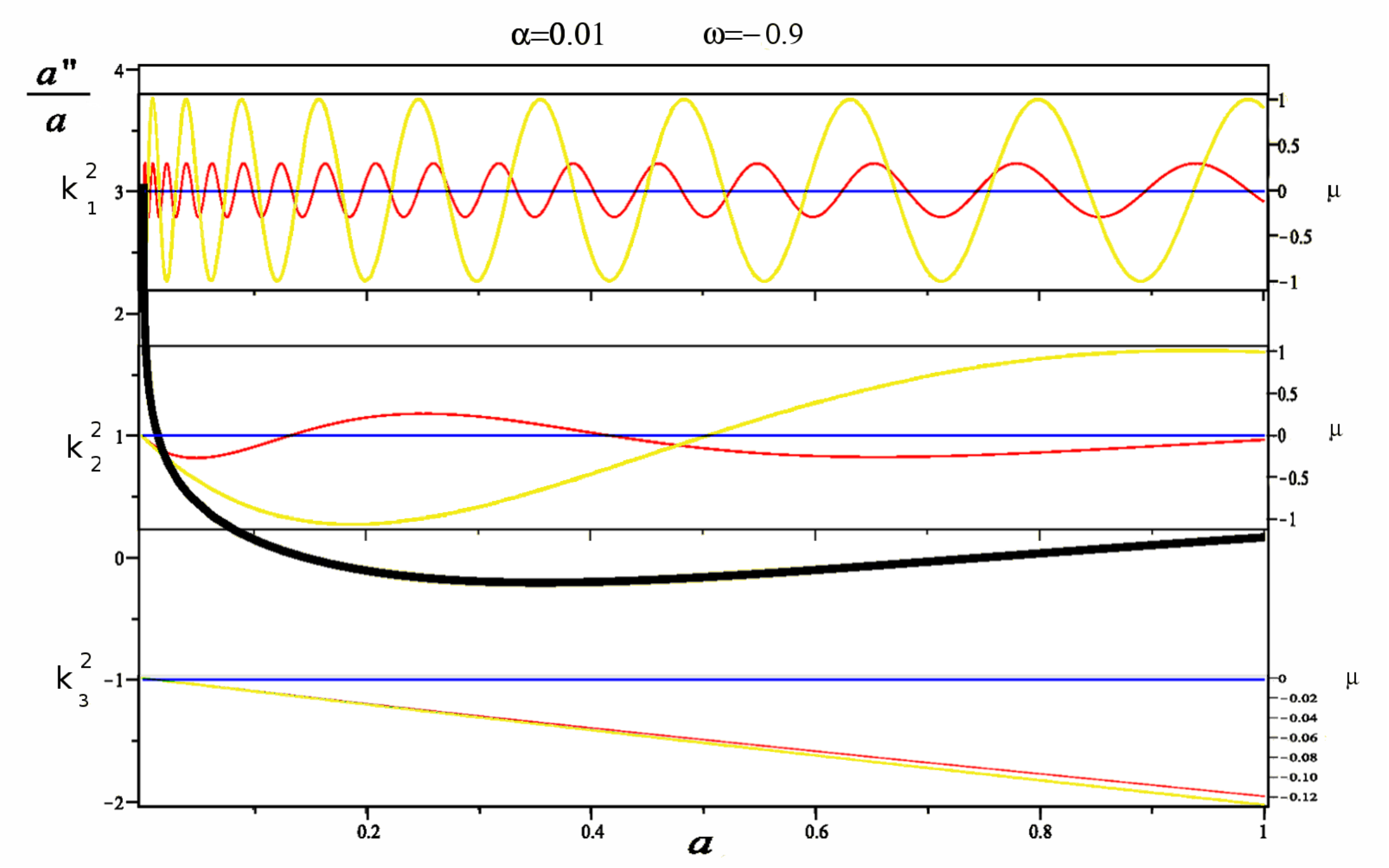}
\includegraphics*[scale=0.4]{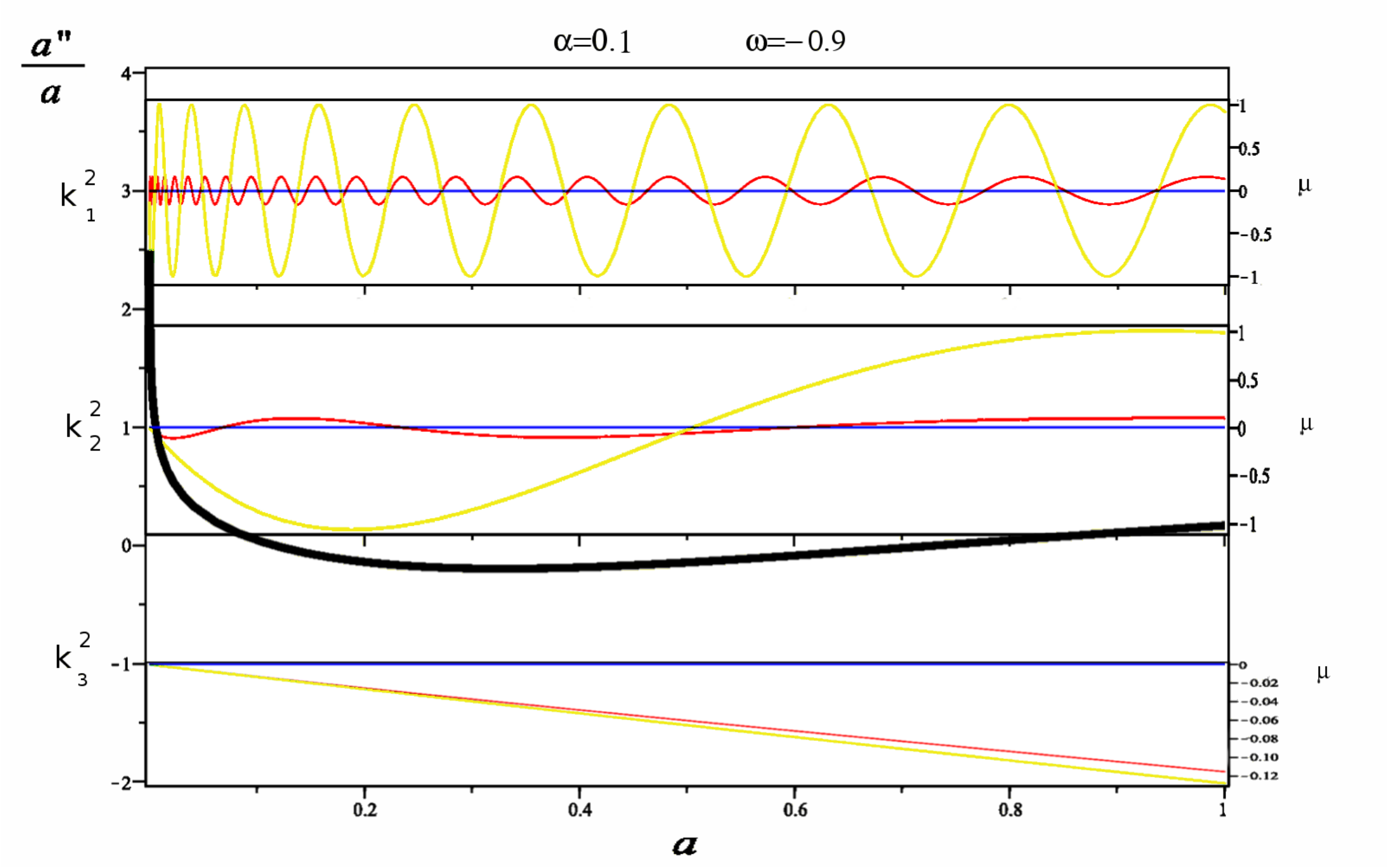}
\caption{RGWs amplitude evolution vs. $a$ in CDE model with $w=-0.9$ and interaction $Q_1$. Refer to section \ref{s4} for details.} \label{fig2}
\end{figure}

\begin{figure}[tbp]
\includegraphics*[scale=0.4]{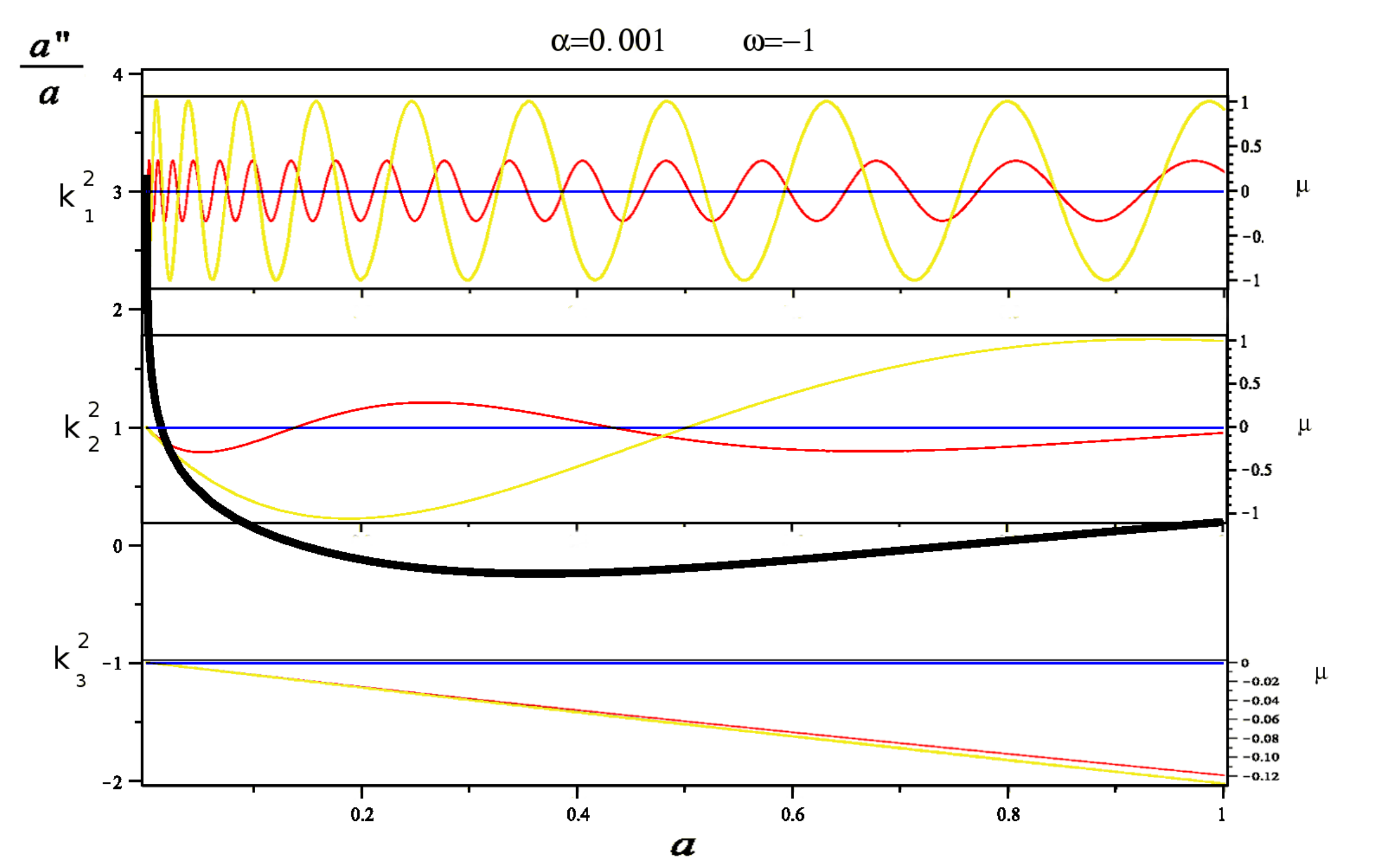}
\includegraphics*[scale=0.4]{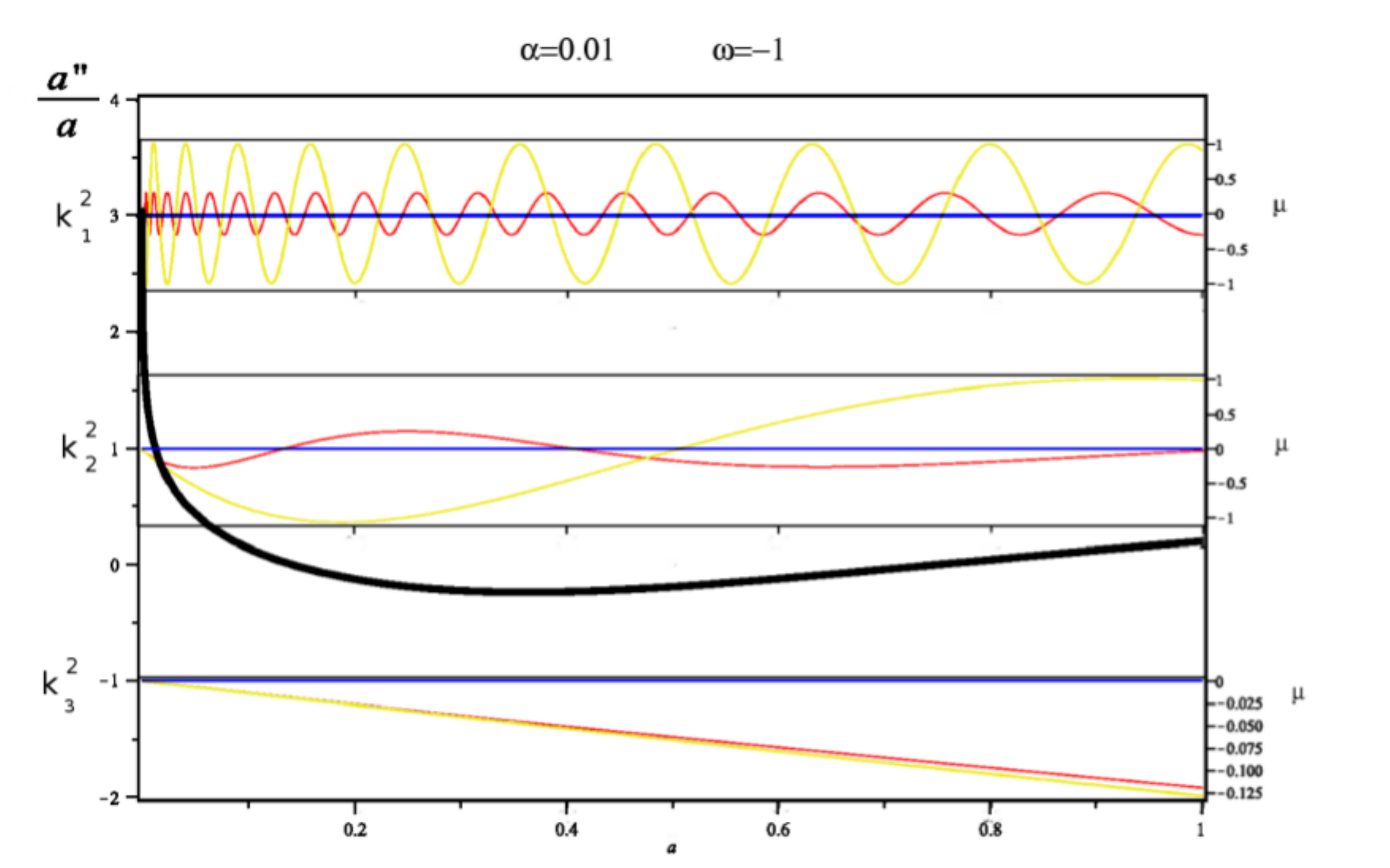}
\includegraphics*[scale=0.4]{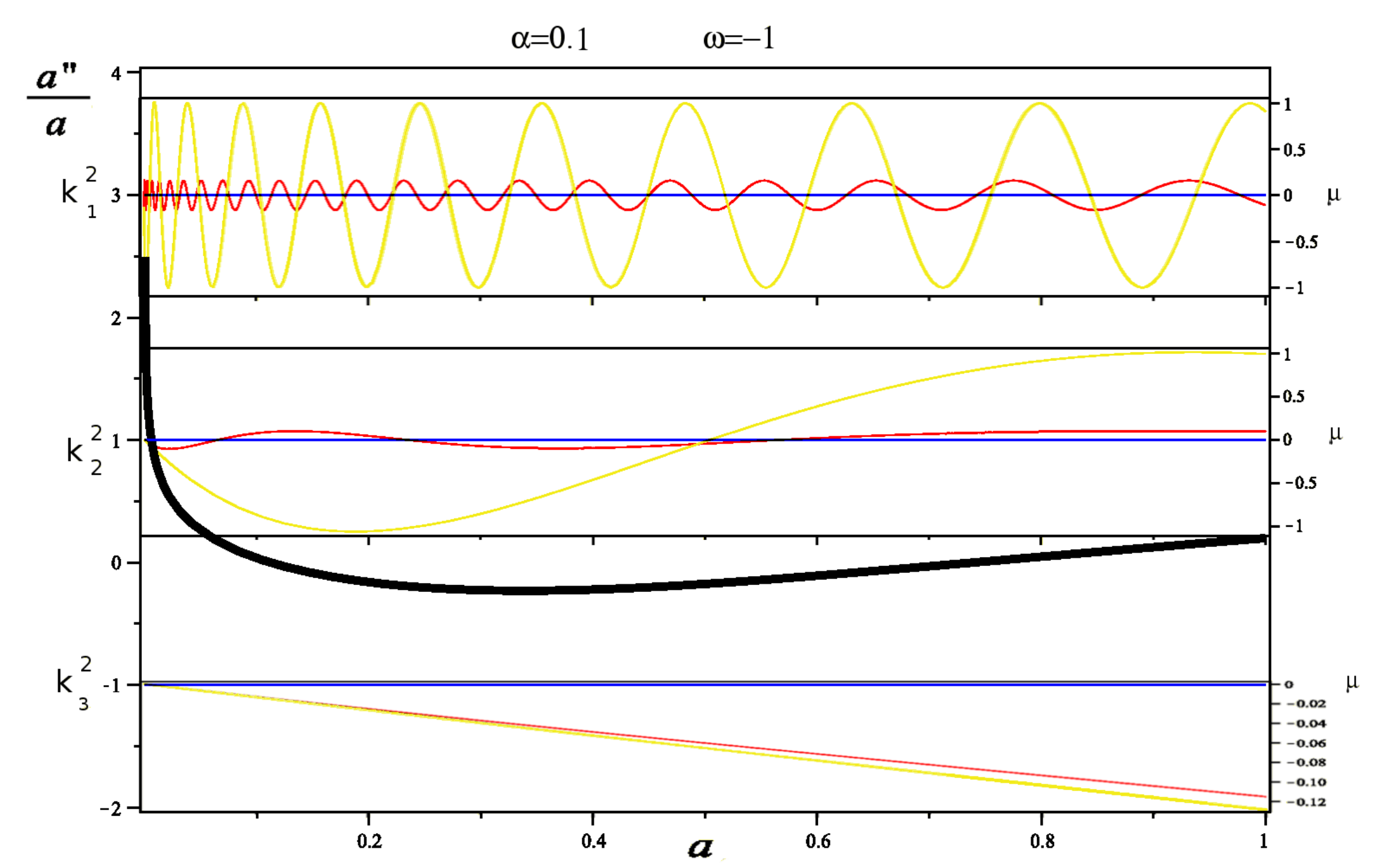}
\caption{RGWs amplitude evolution vs. $a$ in CDE model with $w=-1$ and interaction $Q_1$. Refer to section \ref{s4} for details.} \label{fig3}
\end{figure}

\begin{figure}[tbp]
\includegraphics*[scale=0.4]{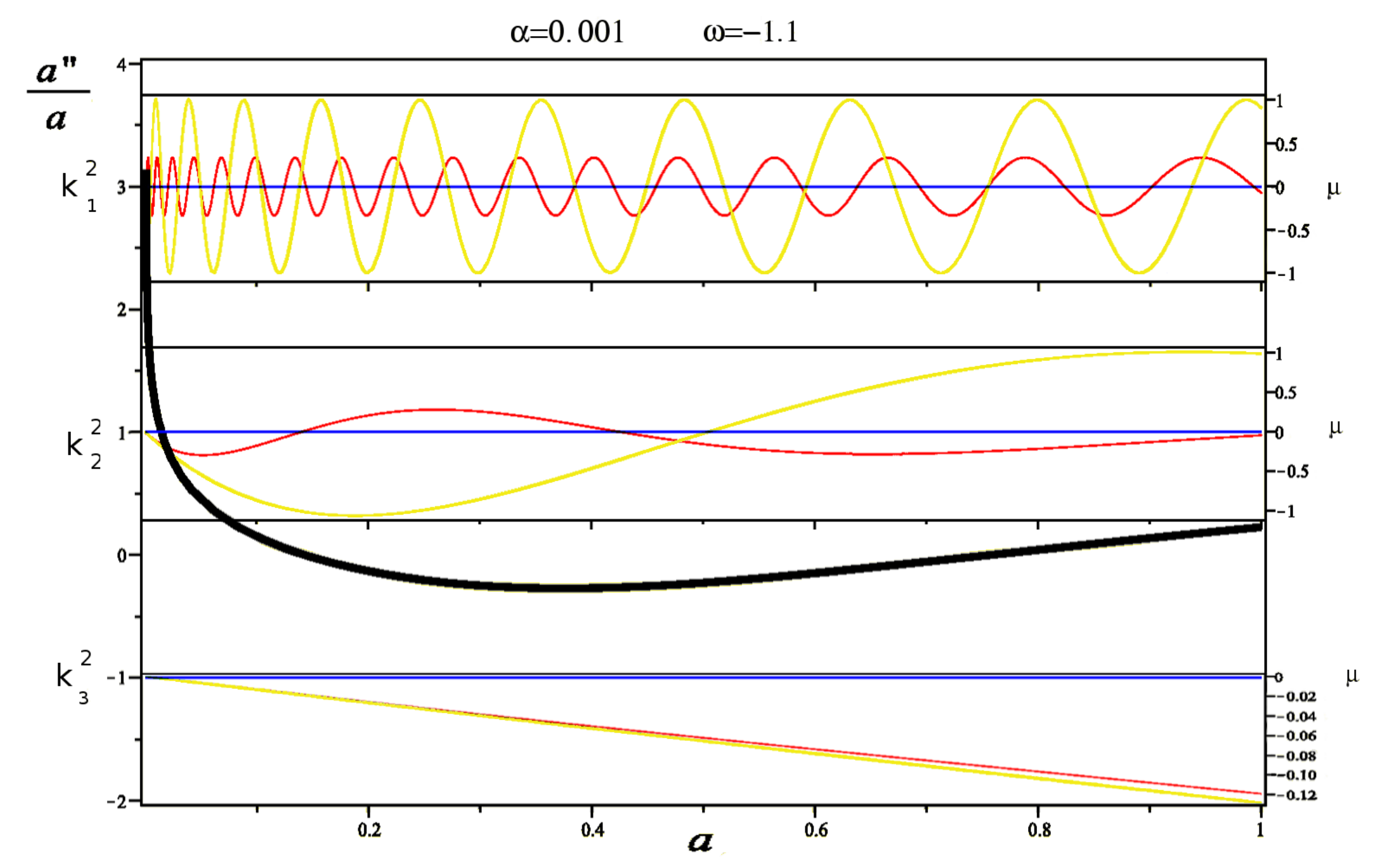}
\includegraphics*[scale=0.4]{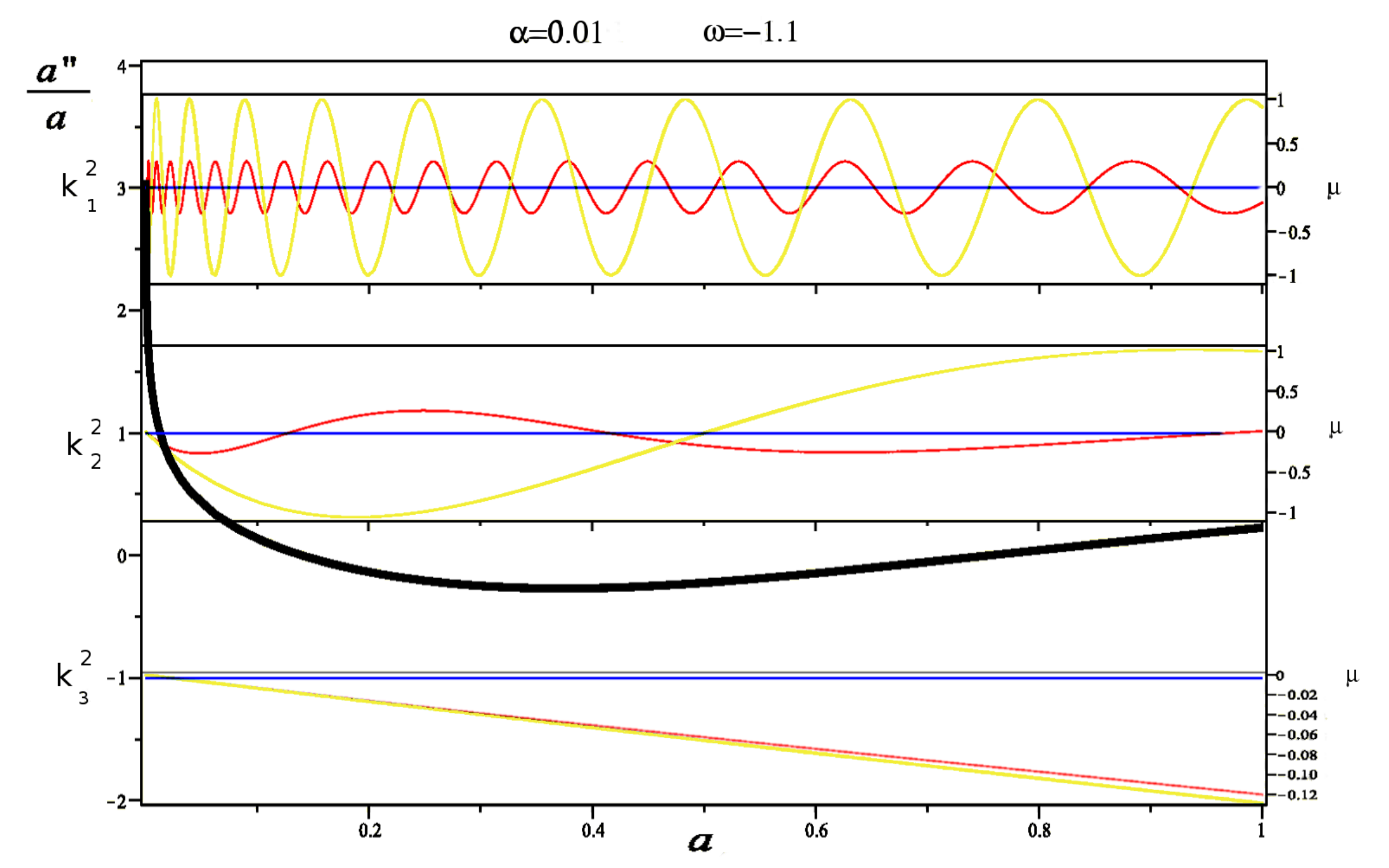}
\includegraphics*[scale=0.4]{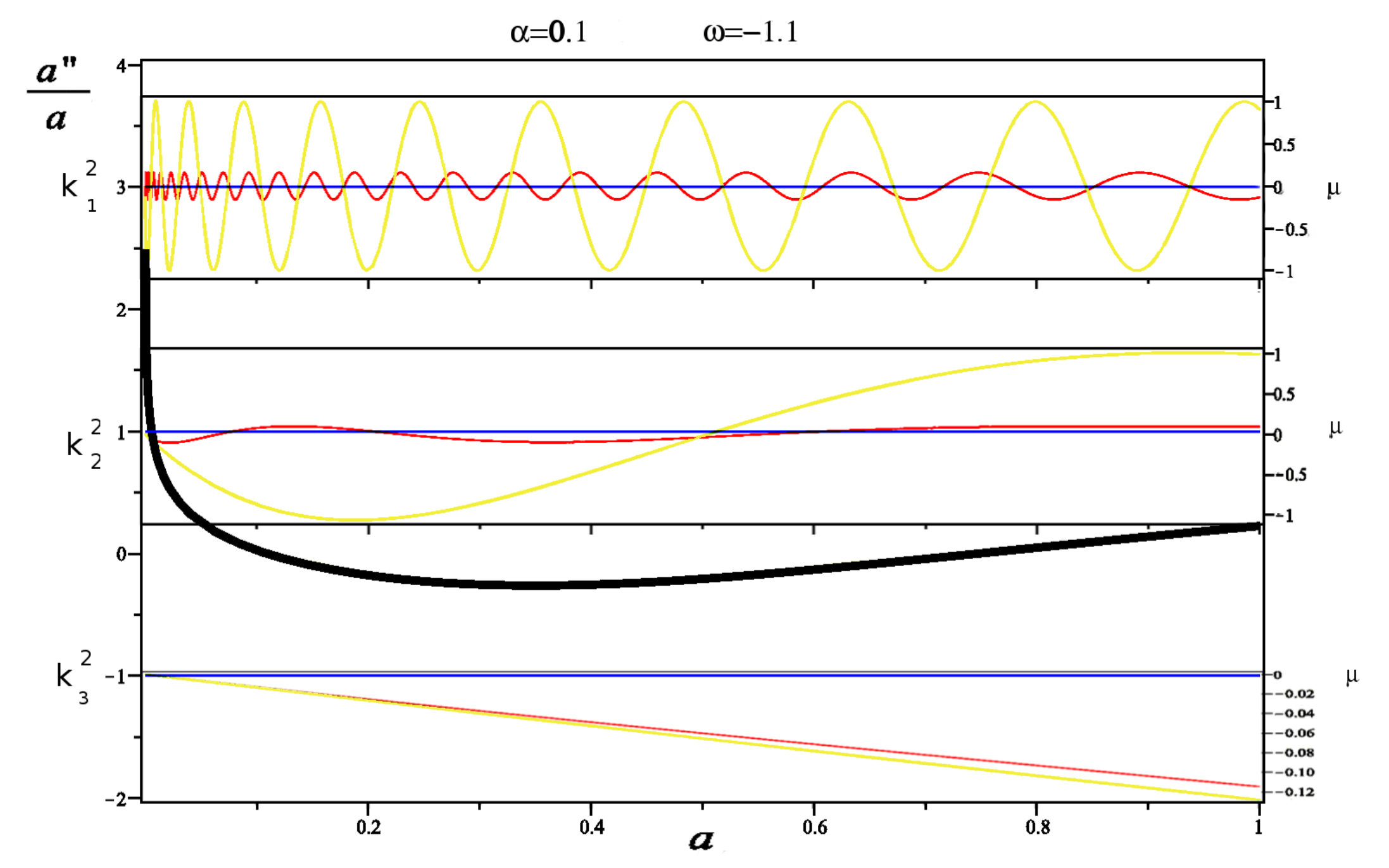}
\caption{RGWs amplitude evolution vs. $a$ in CDE model with $w=-1.1$ and interaction $Q_1$. Refer to section \ref{s4} for details.} \label{fig4}
\end{figure}

\begin{figure}[tbp]
\includegraphics*[scale=0.4]{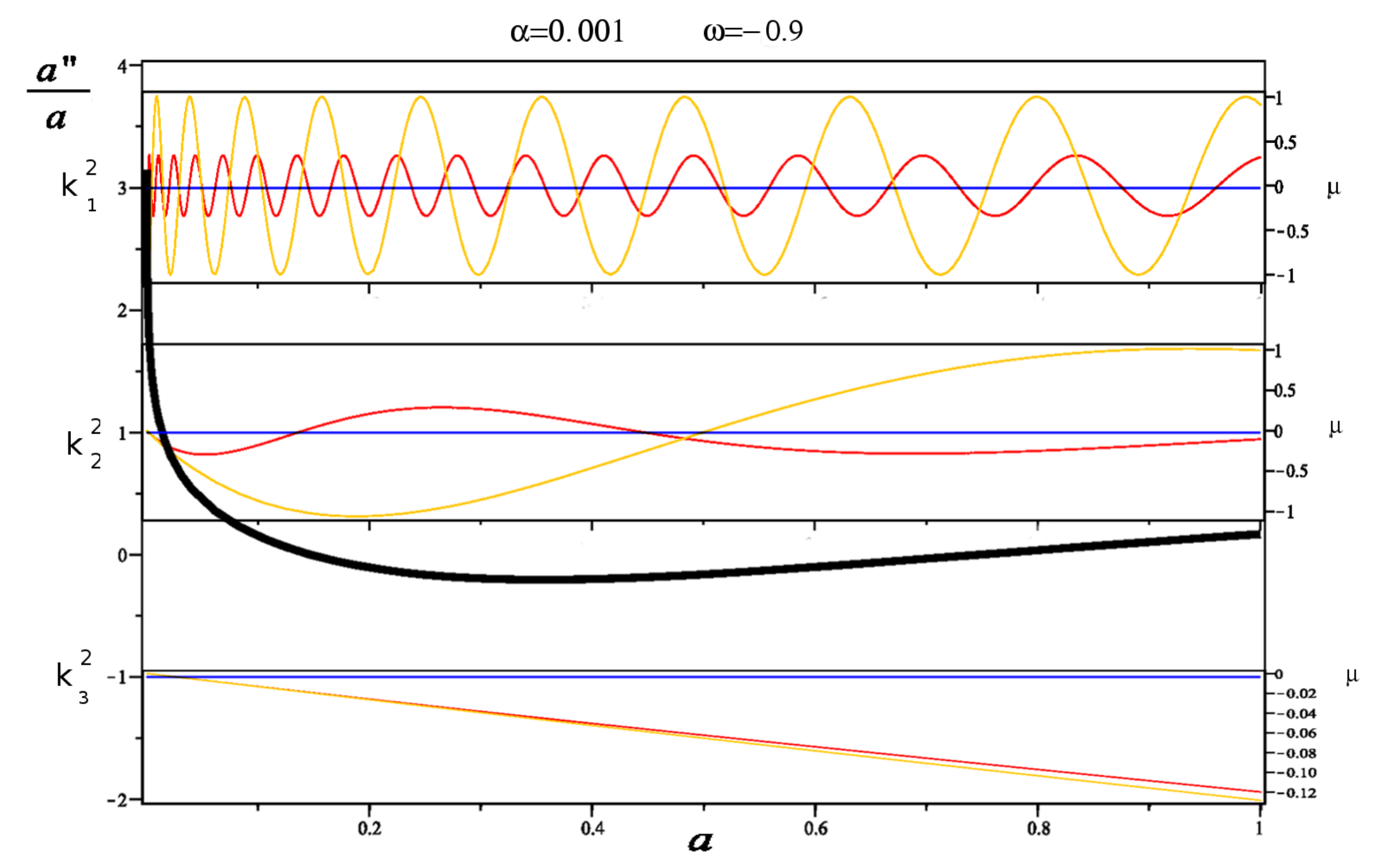}
\includegraphics*[scale=0.4]{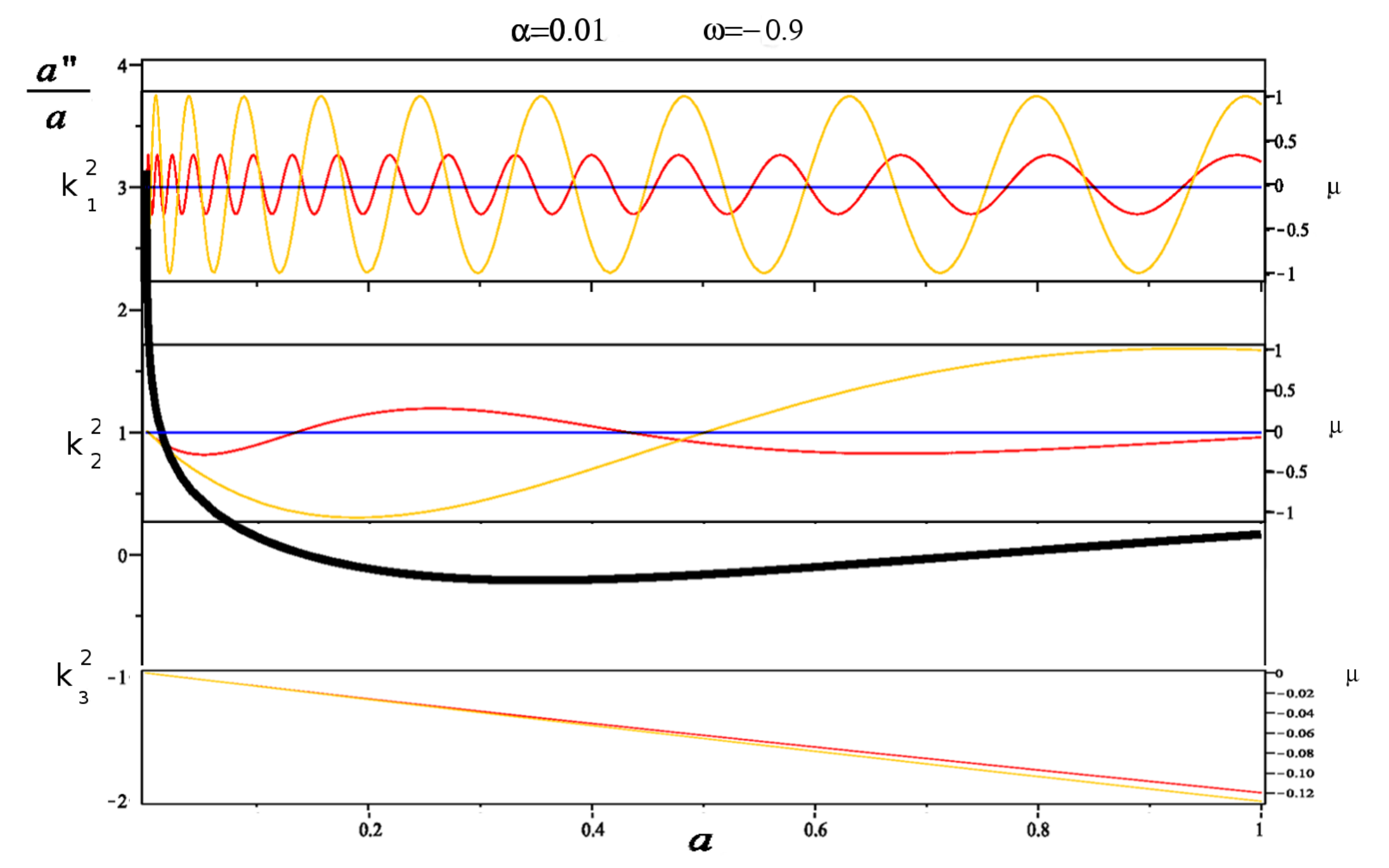}
\includegraphics*[scale=0.4]{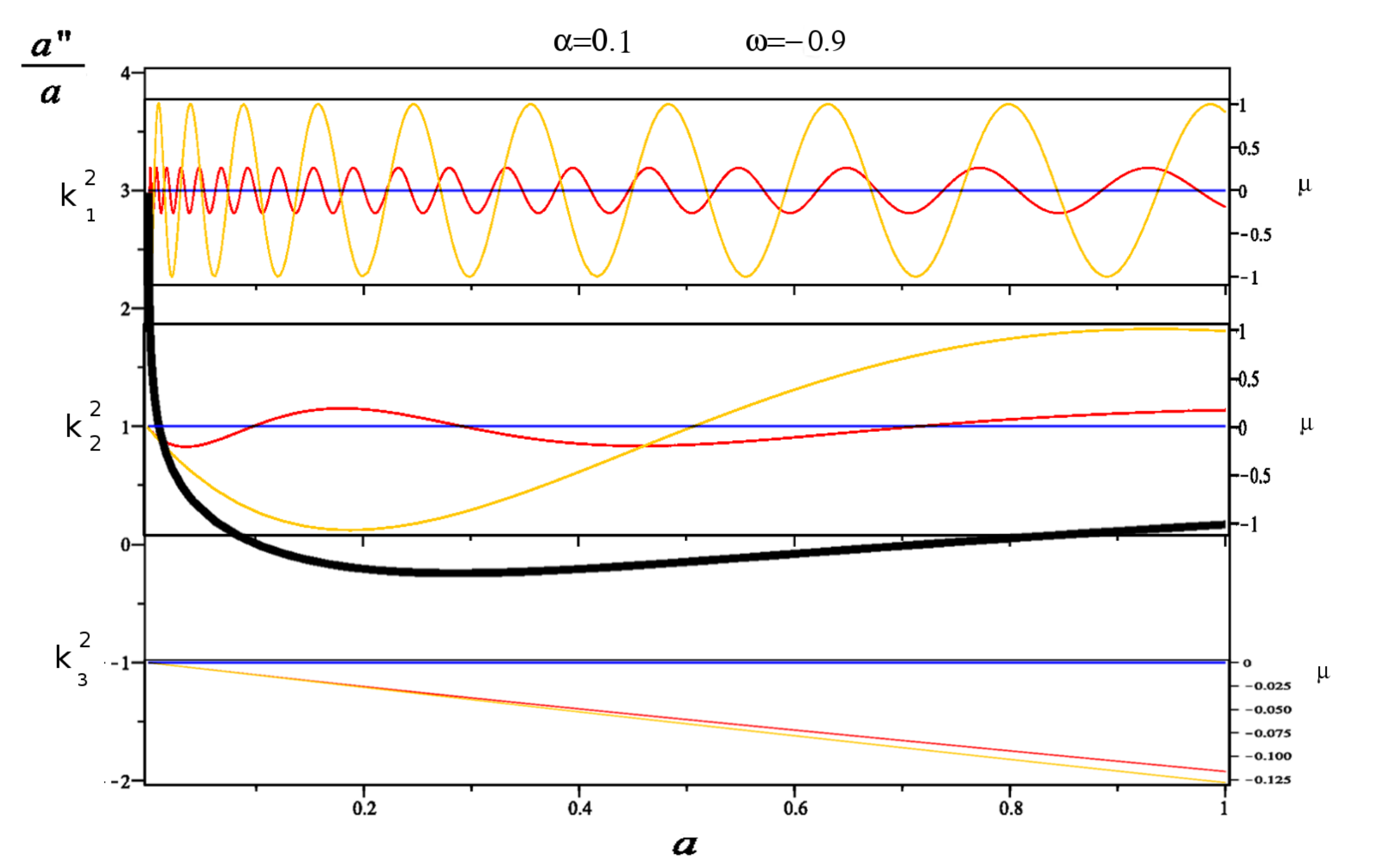}
\caption{RGWs amplitude evolution vs. $a$ in CDE model with $w=-0.9$ and interaction $Q_2$. Refer to section \ref{s4} for details.} \label{fig5}
\end{figure}

\begin{figure}[tbp]
\includegraphics*[scale=0.4]{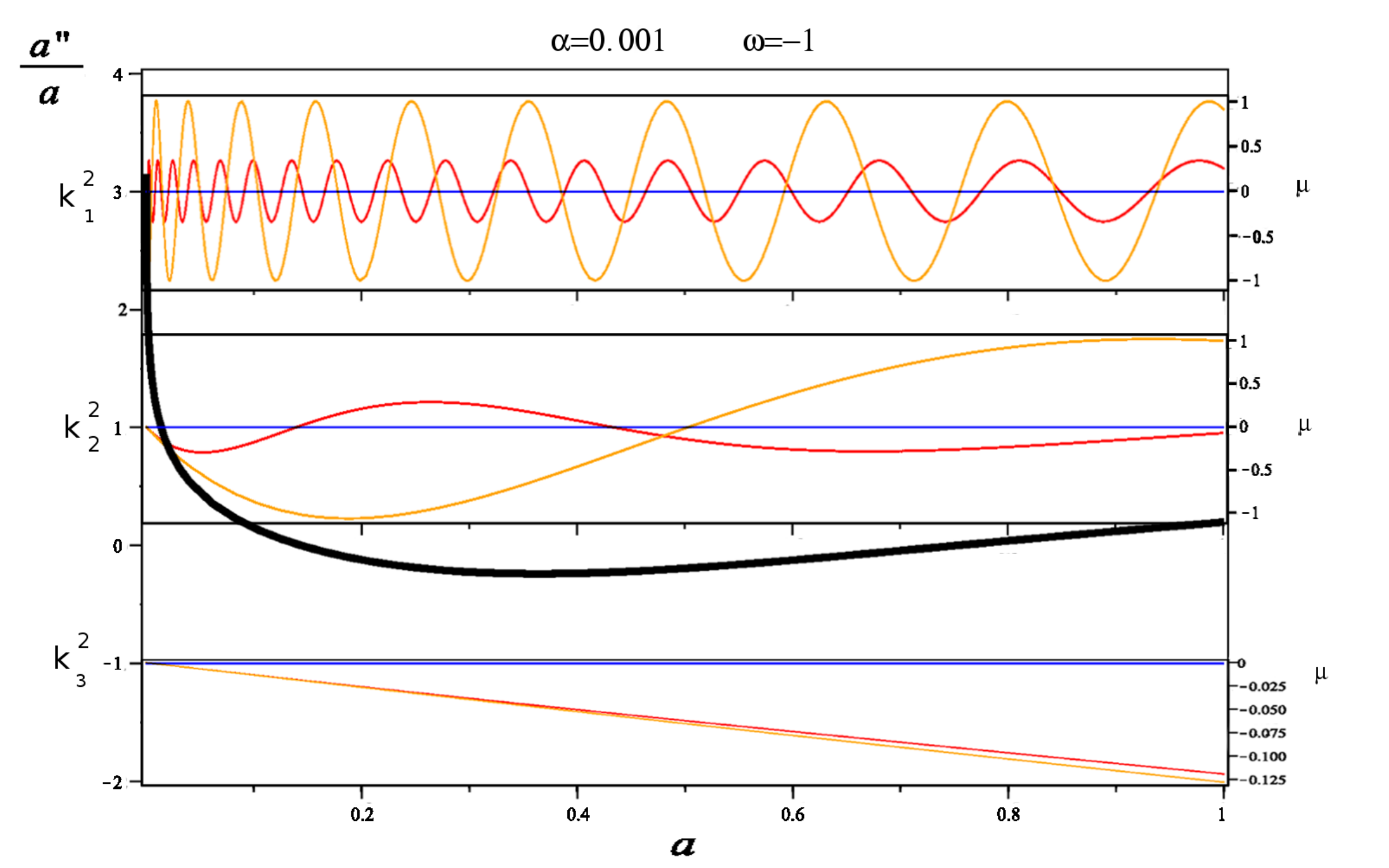}
\includegraphics*[scale=0.4]{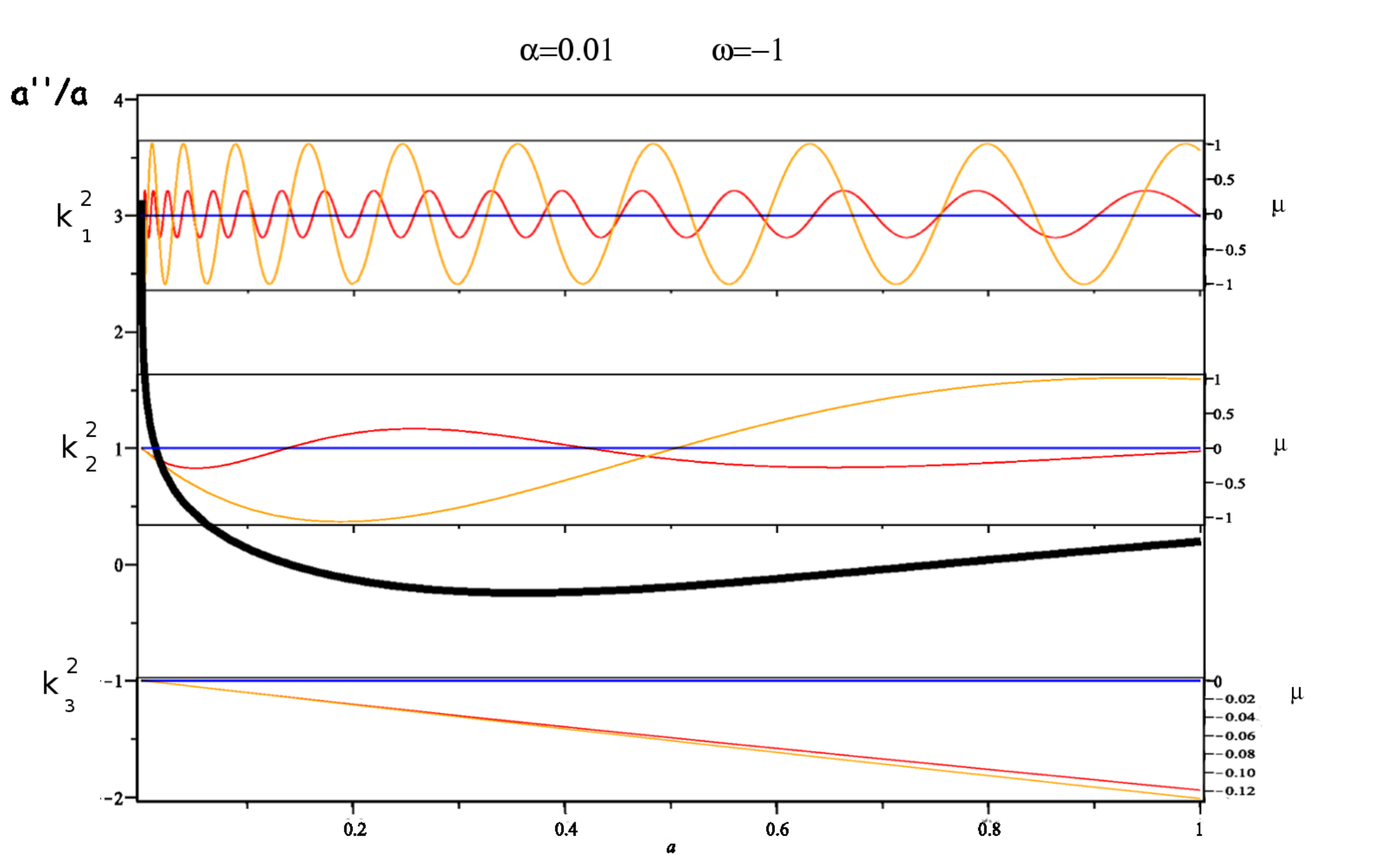}
\includegraphics*[scale=0.4]{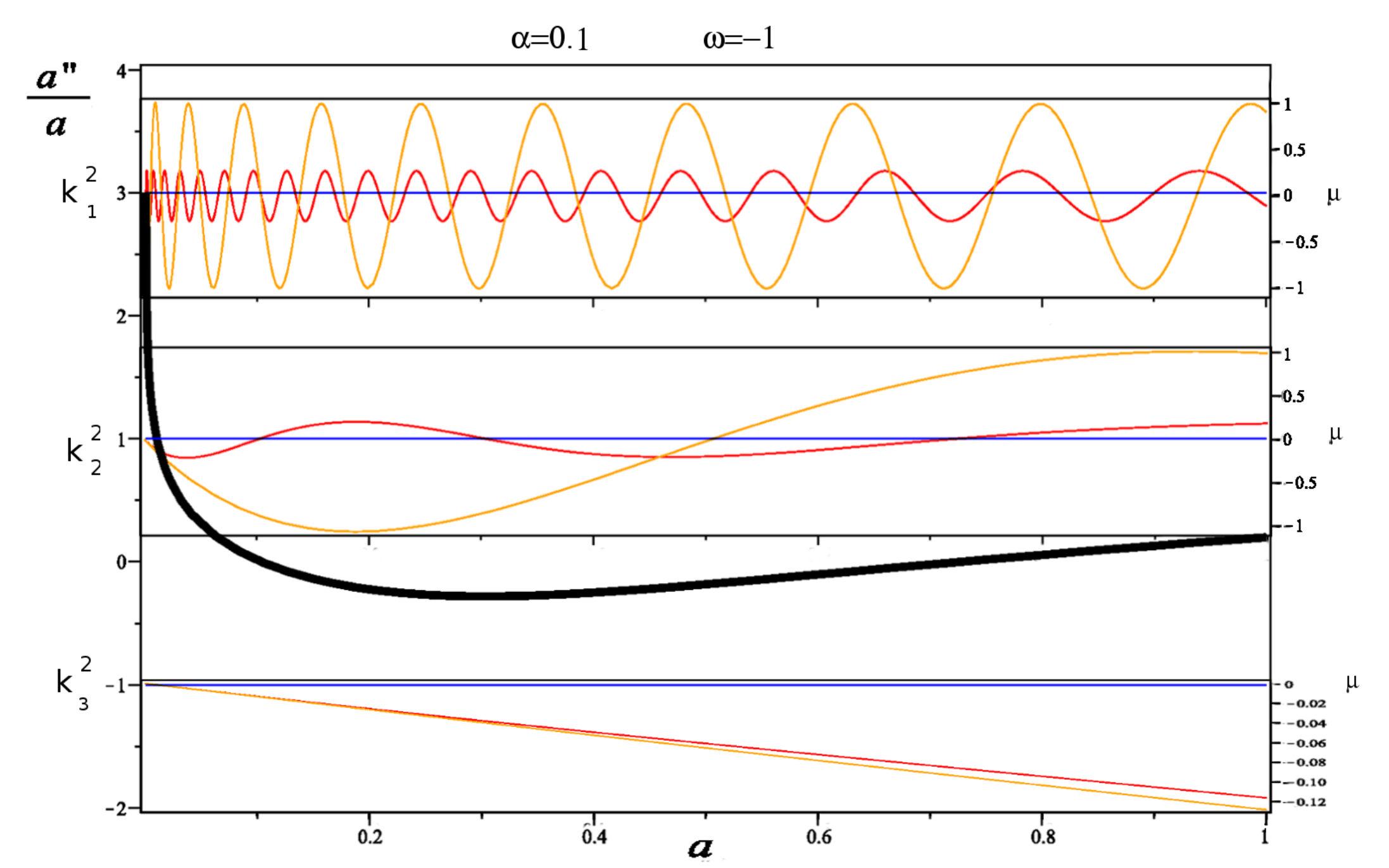}
\caption{RGWs amplitude evolution vs. $a$ in CDE model with $w=-1$ and interaction $Q_2$. Refer to section \ref{s4} for details.} \label{fig6}
\end{figure}

\begin{figure}[tbp]
\includegraphics*[scale=0.4]{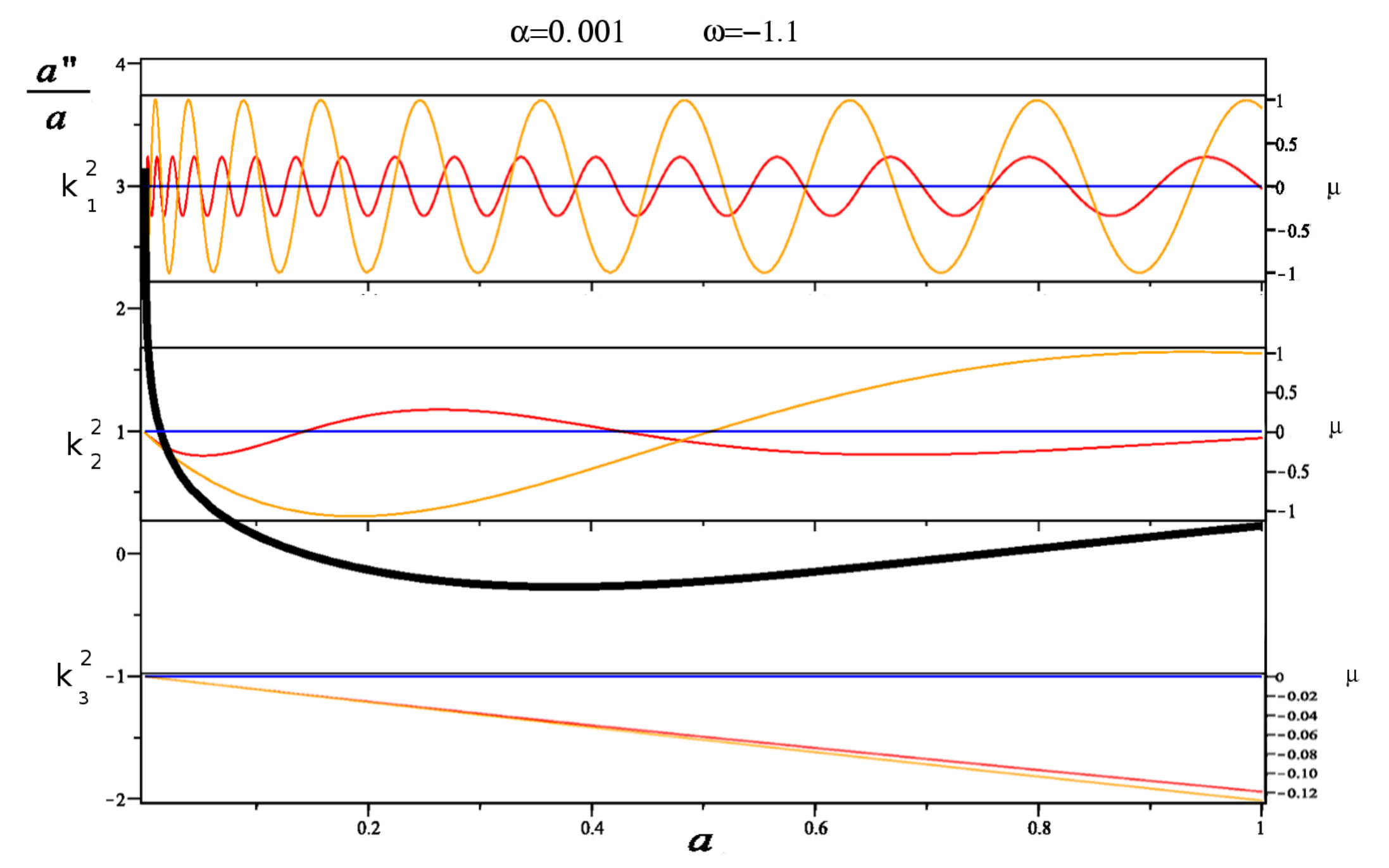}
\includegraphics*[scale=0.4]{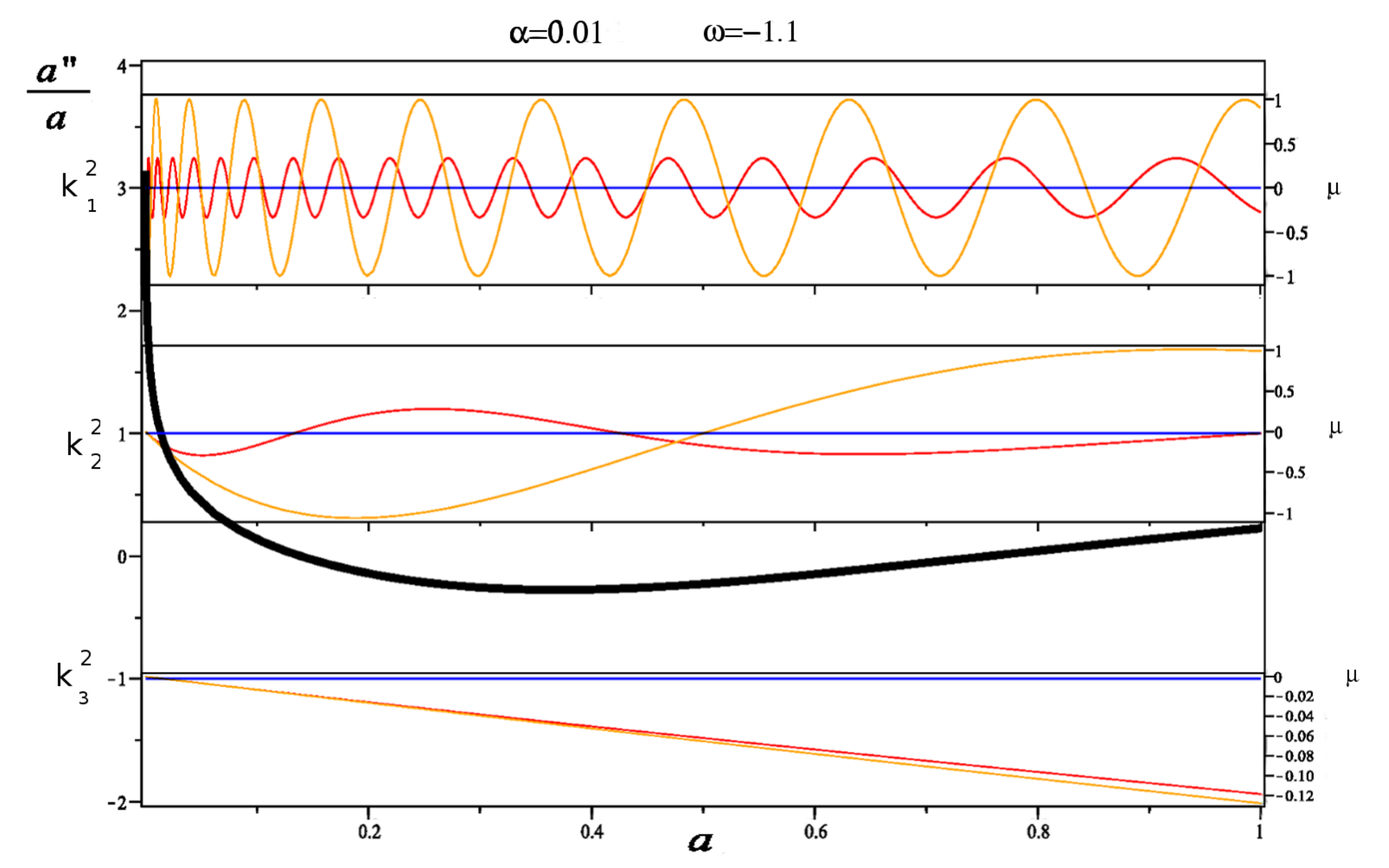}
\includegraphics*[scale=0.4]{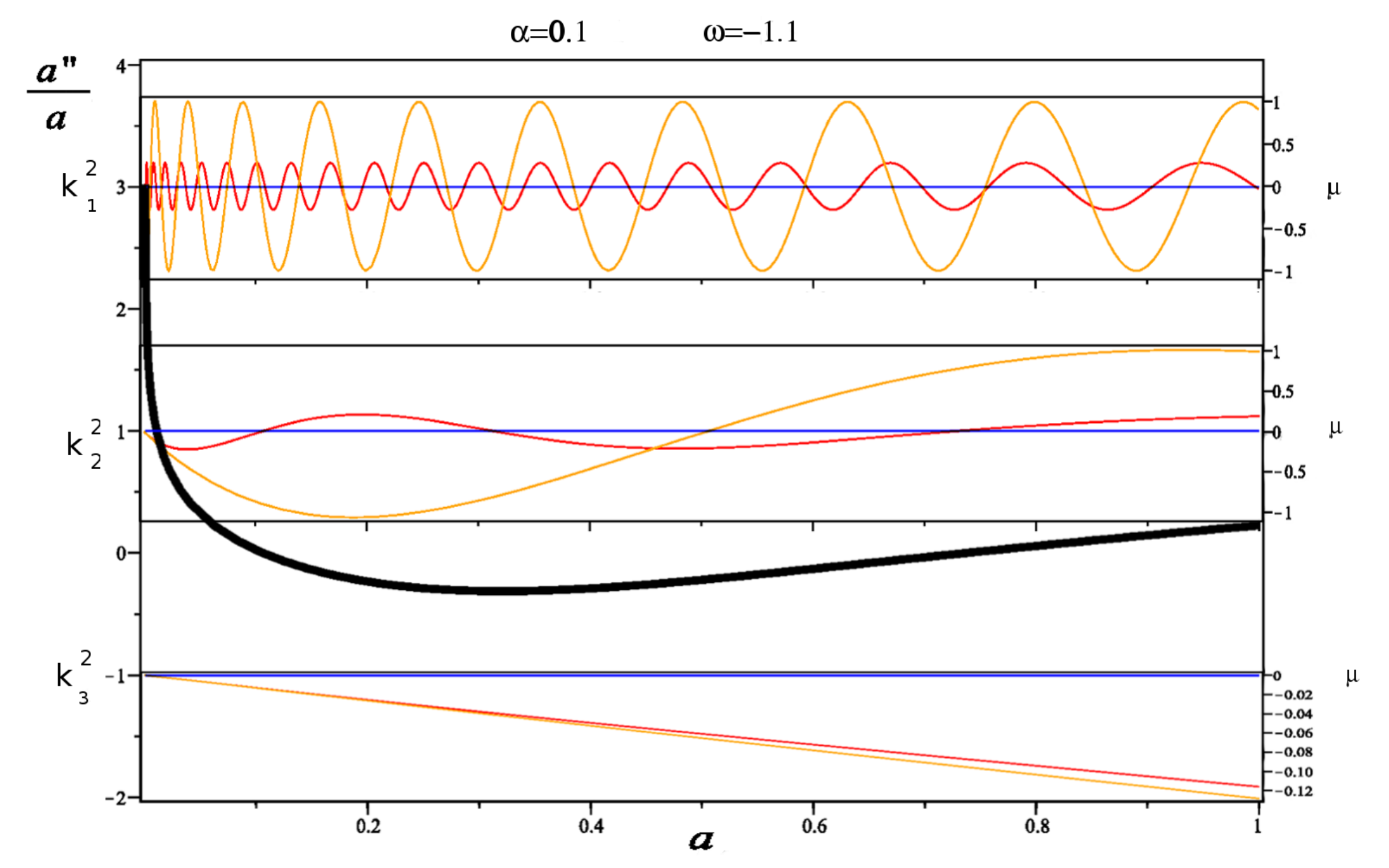}
\caption{RGWs amplitude evolution vs. $a$ in CDE model with $w=-1.1$ and interaction $Q_2$. Refer to section \ref{s4} for details.} \label{fig7}
\end{figure}

The results for interaction $Q_1$ are shown in figures \ref{fig2}-\ref{fig4}, while the interaction $Q_2$ are shown in figures \ref{fig5}-\ref{fig7}.

Some general behavior in the plots can be stated before addressing the specifics of each plot. First, the numerical solutions in the adiabatic regime $\textrm{k}_3=10^{-1/2}$ are almost identical in all plots. This is related with the fact that for those waves, $\mu$ evolves proportional to $a$ (parametric amplification), no matter what choice of the parameters or interaction is made. Second, for the other regimes the numerical solutions $\mu$ have an amplitude of oscillation smaller that the initial $\mu_D$. This is, in part, related to the fact that the potential term decreases faster in terms of $a$ in the scenarios with CDE than the potential in the dust scenario.

Comparing the three panels in figure \ref{fig2} we can state that the bigger the parameter $\alpha$, the smaller amplitude of oscillation is. This fact is related with the decrease of the potential term vs. $a$ in the case of the $\alpha=0.1$ plot. The sooner the RGW leaves the super-adiabatic regime, the smaller the amplitude. For the $\alpha=0.01$ and $\alpha=0.001$ cases, although the potential term is similar, the amplitude is still smaller in the $\alpha=0.01$ plot because of the extra term proportional to $d\mu/da$. This behavior can be observed also in the three panels of figures \ref{fig3} and \ref{fig4}, respectively.

In \cite{ol}, the authors report that CDE models present a damping in the power spectrum of density perturbations compared to non-interacting DE models. In this work we also find a damping in the amplitude of RGWs for CDE models. In fact, evolution of RGWs and of density perturbations are ruled by similar equations and mechanisms (the former being scalar perturbations of the background metric, the latter being  second order tensorial perturbations). In both mechanism, the scale factor defines the amplification experimented by the perturbations. In this sense, our results  agree and complement the results reported in \cite{ol}.

Comparing the first panels of figure \ref{fig2}, figure \ref{fig3} and figure \ref{fig4} we can state that for the same $\alpha$ and $\textrm{k}$, the choice of $w$ has an influence in the frequencies of oscillation of the RGWs at a larger scales of $a$, but they present almost the same amplitudes.  Comparing second and third panels of the figures, we can appreciate the same behavior.

Similar conclusions can be asserted from the numerical plots of the interaction $Q_2$. The bigger the $\alpha$ parameter, the smaller the amplitude of RGWs. And comparing the first panels of figures \ref{fig5}-\ref{fig7}, we can state that $w$ parameter has an impact on the frequency of the waves at late values of $a$, but there is no appreciable change in amplitudes.

Finally, we can compare the plots of the two interactions for the same choice of parameters $\alpha$, $w$ and $\textrm{k}$. For $\alpha=0.001$, we do not appreciate significant differences in amplitudes of $\mu$ and frequencies are also very similar. A slight difference can be appreciated when comparing $\alpha=0.01$ plots of $Q_1$ with those of $Q_2$, $Q_1$ solutions presenting a smaller amplitude than $Q_2$ solutions. $\alpha=0.1$ plots for $Q_1$ present a significative smaller amplitudes that those of interaction $Q_2$. The frequencies are also different in both cases. From those results, we can conclude that interaction $Q_1$ generates RGWs with smaller amplitudes than interaction $Q_2$. This fact can be explained by the evolution of CDM density in both interactions. Interaction $Q_1$ predicts less CDM present at the beginning of the dust era than interaction $Q_2$.

\section{Power spectrum}
\label{s5}
In this section we compute the power spectrum of RGW for the CDE models. For the sake of clarity, we recover units for $H_0$ and consider the WMAP value $H_0=70 km/s/Mpc \simeq 2.27 \times 10^{-18} s^{-1}$ \cite{WMAP}. Note that the wave number $k$ is not a physical quantity (it is defined as a commoving quantity), the corresponding physical frequency is defined as $\omega (a) = k/a$. Given that $a_0=1$, we can state that $\omega_0=k$, i.e. the frequency of the waves observed today corresponds to the wavenumber $k$. In this section, we will refer indistinctly to the present day frequencies and the wave numbers of the RGW $k$.

As we have stated in section \ref{s2}, waves whose wave number $k^{2}\gg \frac{a^{\prime\prime }}{a}$ evolve as free waves and are not affected by the dynamics of the universe. For $a=a_2=10^{-4}$, we define the bound $K_2$ as
\be
K_2^2=\frac{a''}{a}(a_2)=2a_2^2 H^2(a_2)+a_2^3 H(a_2) \frac{dH}{da}(a_2),
\ee
which depends on parameters $\alpha$ and $w$ through $H(a_2)$ and $\frac{dH}{da}(a_2)$. RGW with $k \gg K_2$ are not affected by the CDE potential, and their amplitude is proportional to

\[
R(k)=\frac{C_{D}}{C_{I}}\sim \left\{
\begin{array}{c}
1\qquad (k\gg -1/\eta _{1}=a_1 H_1), \\
k^{-2}\qquad (a_1 H_1\gg k\gg K_2), \\
\end{array}%
\right.
\]
where $a_1$ and $H_1$ are the scale and Hubble factors at the end of the inflation, respectively. Although those values depend on the inflationary model considered we will assume the typical slow roll inflation value $H_1= 10^{35} s^-1$. It is straightforward that
\[
a_1=10^{-4}\left(H(10^{-4})/H_1\right)^{1/2} \sim 10^{-24},
\]
with $H(10^{-4})$ being the Hubble factor at the beginning of the CDE era, which depends on $\alpha$ and $w$ but in any case is of order $10^{-5} s^{-1}$.

The waves with $k\gg a_1 H_1\sim 10^{11} s^{-1}$ did not experiment any adiabatic amplification and have at the present day an amplitude several orders of magnitude smaller than the amplitude at the instant they were generated. Consequently, those waves do not contribute to the power spectrum of RGWs.

Waves with $a_1 H_1\gg k\gg K_2$ were amplified adiabatically during the inflation and resumed the evolution as free waves after it. RGW with $k \ll K_2$, experimented a an amplification during inflation and a second one in the CDE era. Both set of RGW have to be considered in the power spectrum.

On the other hand, the perturbations whose wave number is $k\ll (a^{\prime \prime }/a)(a=1)$ have wavelengths larger than the Hubble radius of the universe, i.e. in the whole history of the universe, they have not completed one single period of oscillation. Those perturbation cannot be considered waves. This fact puts lower bound on the wave number for the RGW spectrum, $K_0$ where
\be
K_0^2=\frac{a''}{a}(a=1)=2H^2_0+ H_0 \frac{dH}{da}(a=1).
\ee
The term $\frac{dH}{da}$ of the above equation depends on $\alpha$ and $w$, but in all the cases considered $K_0$ is of order $H_0$.

We consider the power spectrum defined as in \cite{iz03}, $P(k)=k h(k)^2$ where $h(k)=|\mu (k)|/a$ and $|\mu (k)|$ is the amplitude of oscillation of function $\mu$. The waves with $K_2 \ll k \ll K_1$ have the same power spectrum as in \cite{grish74, grish93, iz04, iz03}
\be
P(k)=\frac{\hbar }{4\pi ^{2}c^{3}}a_{1}^{4}H_{1}^{4}k ^{-1}.
\ee
The power spectrum of the waves with $K_0 \ll k \ll K_2$ is obtained by numerically computing $\mu(a)$ for each $k$ as in the previous section, and finally choosing the maximum value of $\mu(a)$ in the last period of oscillation as $|\mu (k)|$.

\begin{figure}[tbp]
\includegraphics*[scale=0.3]{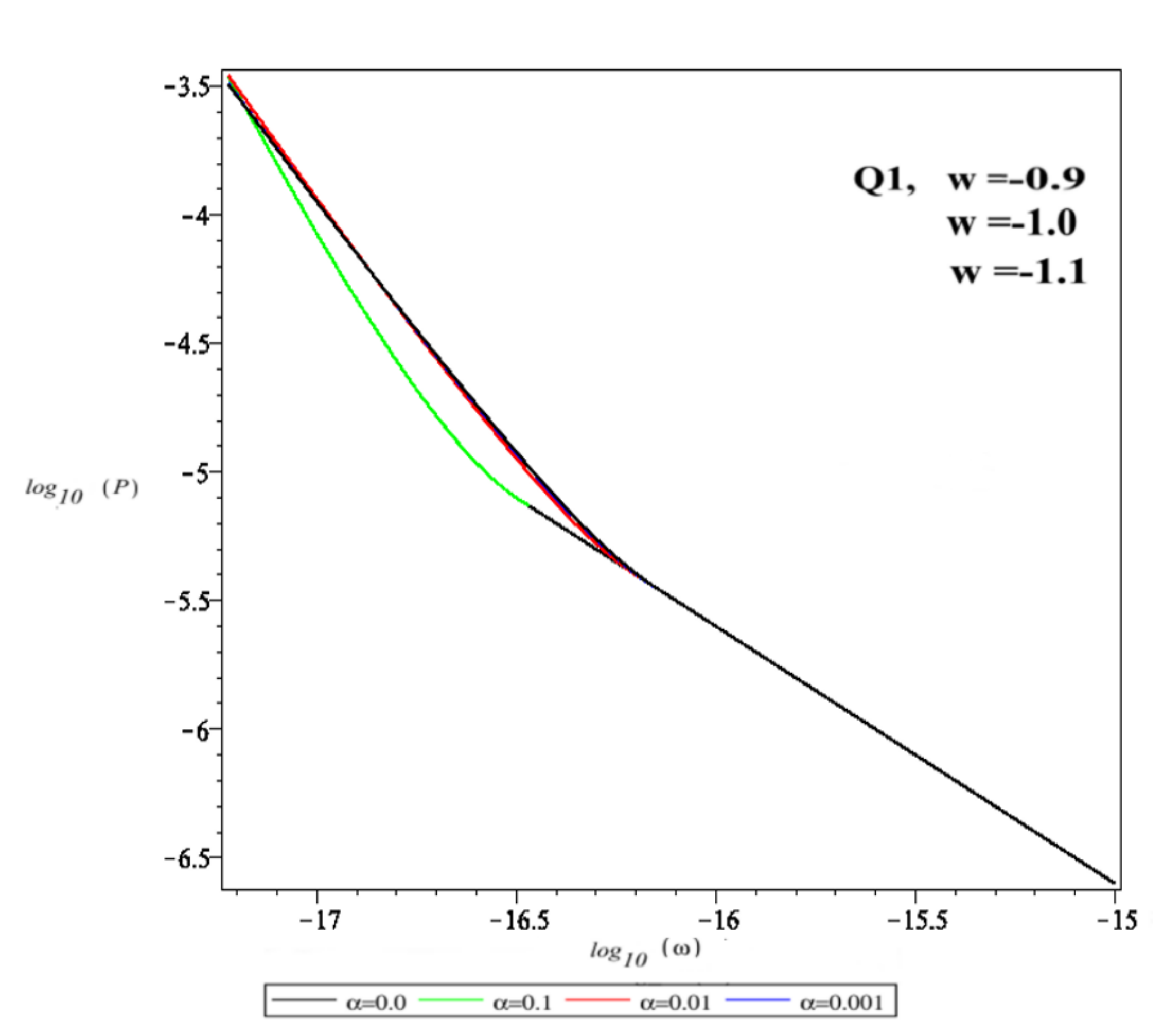}
\includegraphics*[scale=0.3]{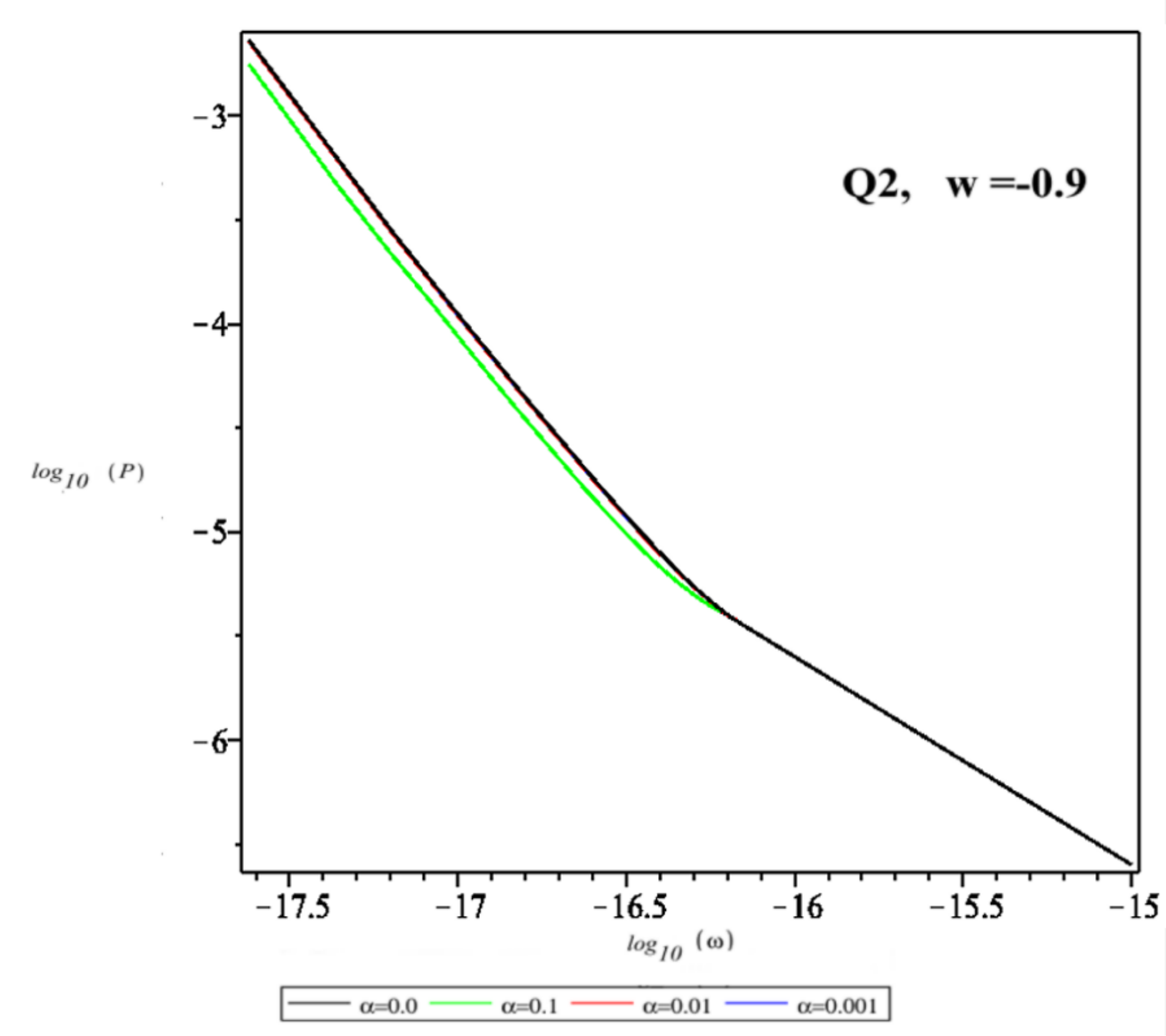}
\includegraphics*[scale=0.3]{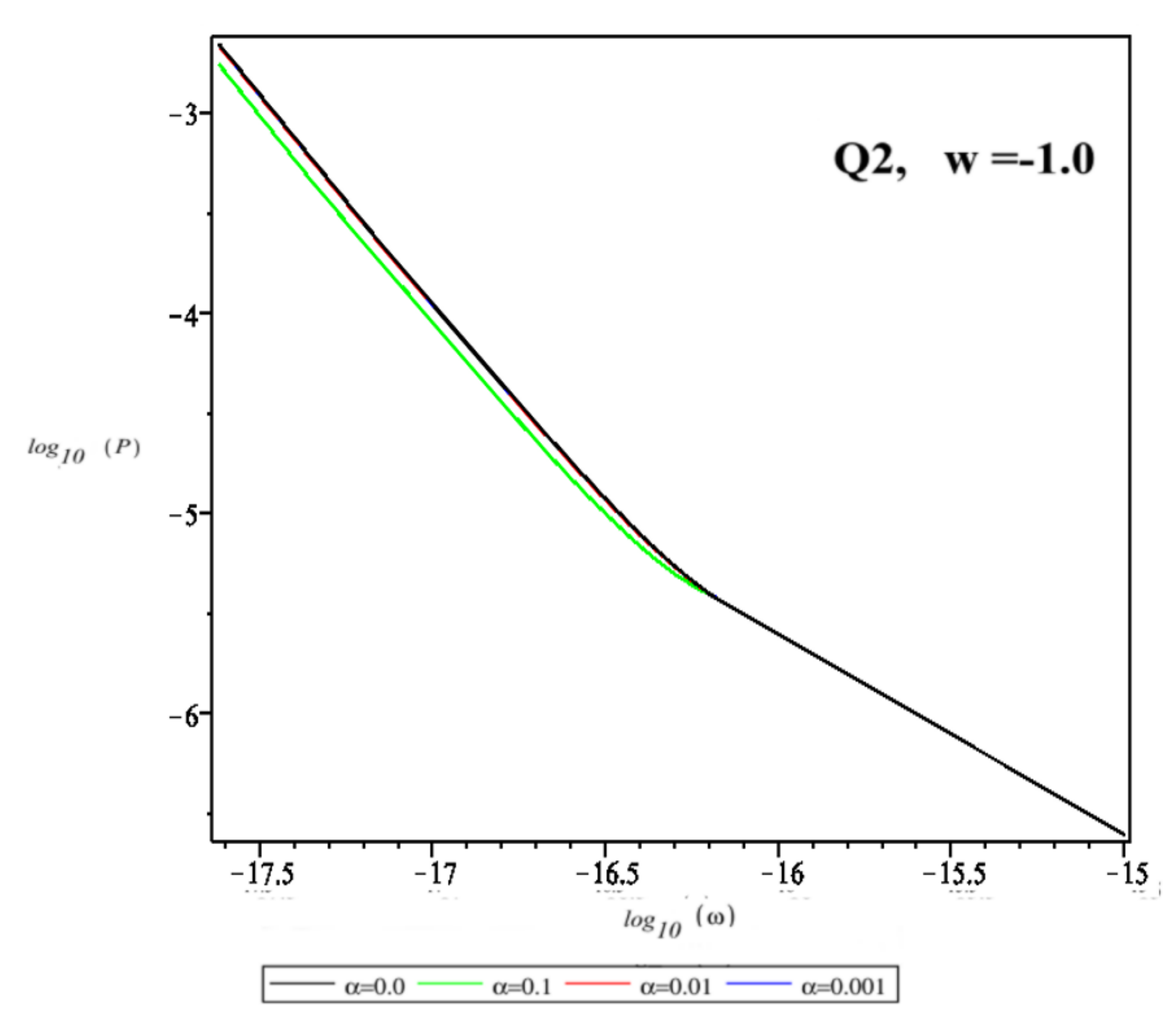}
\includegraphics*[scale=0.3]{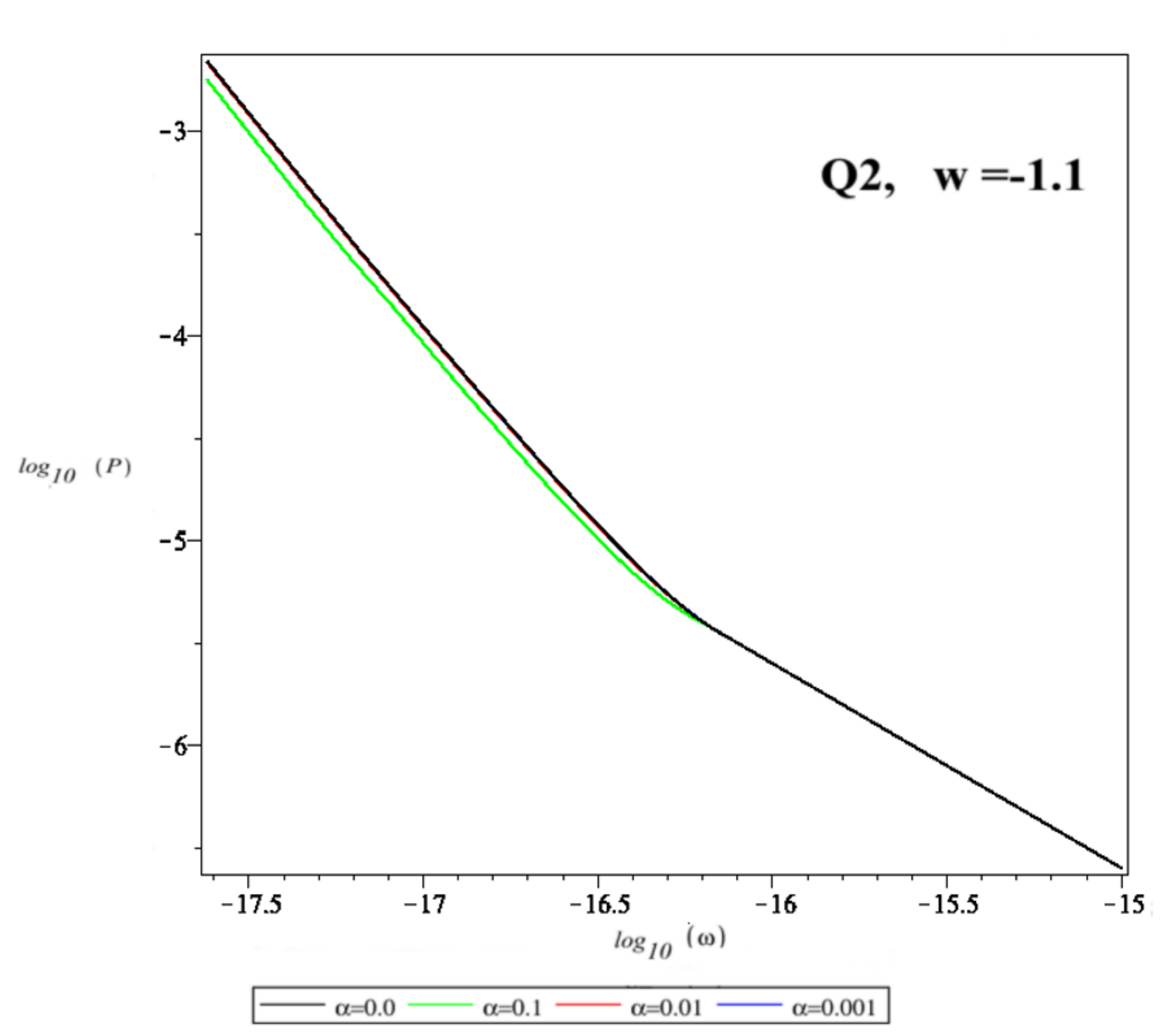}
\caption{Logarithm of RGWs power spectrum vs. logarithm of frequency in CDE model for interactions $Q_1$ and $Q_2$ and different choices of the parameters $w$ and $\alpha$. The power spectrum $P$ and the frequency $w$ are expressed in $erg \cdot s/cm^3$ and $s^{-1}$, respectively} \label{fig8}
\end{figure}

Figure \ref{fig8} shows $log_{10}(P(\omega))$ vs $log_{10}(\omega)$ of the RGWs in the CDE model with both interactions, and some choices of parameters $\alpha$ and $w$. For the interaction $Q_1$, the power spectrum of the $w=-0.9,-1,-1.1$ cases are very similar and the logarithmic plots are almost the same. Second to forth panels correspond to interaction $Q_2$, for which the choice of parameter $w$ has an impact on the power spectrum. When the interaction parameter is $\alpha=0.001$, the resulting power spectrum is almost identical to the non-coupled model $\alpha=0.0$ for both interactions considered.

\section{Conclusion}
\label{s6}

The evolution of RGWs is determined by the scale factor of the LFRW metric. CDE models predict a scale factor evolution different from a non-coupled dark energy models. While in the latter models, a sudden transition approximation between dust and DE stages can be assumed in order to solve the wave equation, in the former models a smooth transition between dust stage and present day follows. For some choices of the interaction term, the CDE stage evolve to an attractor of the system, where $\rho_c/\rho_{\phi}$ tends to a constant value solving the coincidence problem. For CDE scenarios, a numerical solution to the equation for the RGWs evolution is needed.

In this work, we assume two different coupling terms of CDE: $Q_1=\alpha H \rho_{\phi}$ and $Q_2=\alpha H \rho_{c}$. We solve numerically the equation for the wave in terms of the scale factor $a$ instead of the conformal time $\eta$, for different choices of the coupling parameter $\alpha$, the adiabatic constant of the dark energy $w$, and different wave numbers $\textrm{k}$.

We find that the larger the interaction parameter $\alpha$, the lower the amplitude of the RGWs, in both interactions considered. This result agrees with previous results found in the literature for density perturbations evolution in CDE models \cite{ol}. The mechanism of amplification of RGWs is similar to that of the density perturbation, both being perturbations of the background LFRW metric.

We also find that for the same choice of parameter $\alpha$ and $Q$, the adiabatic coefficient $w$ has an impact on the frequency of the wave for values of $a$ near the present day $a_0=1$,  but almost the same amplitude.

Finally, when comparing both interactions for the same choice of parameters, we find that interaction $Q_1$ predicts a smaller amplitude than interaction $Q_2$.

The amplitudes of the RGWs contribute to the power spectrum of the RGWs. In the hypothetic case that the RGWs power spectrum is determined (or bounded) by observational data, we will obtain a criterium to validate or discard different dark energy models (coupled or not) through RGW evolution. Given that dark energy does not interact with ordinary matter as far as we know, those indirect measurements would be very valuable.

\end{document}